\documentclass[showpacs,showkeys,preprintnumbers,amsmath,amssymb,twocolumn,pra,floatfix]
{revtex4}

\usepackage{amssymb,amsfonts}
\usepackage{amsmath}
\usepackage[dvips]{graphicx}
\begin{document}

\title{Modeling of Low and High Frequency Noise by Slow and Fast
Fluctuators}

\author{Alexander I. Nesterov}%
   \email{nesterov@cencar.udg.mx}
\affiliation{Departamento de F{\'\i}sica, CUCEI, Universidad de Guadalajara,
Av. Revoluci\'on 1500, Guadalajara, CP 44420, Jalisco, M\'exico}

\affiliation{Theoretical Division and the CNLS, MS-258, Los Alamos National Laboratory, Los
Alamos, NM 87544, USA}

\author{Gennady P.  Berman}
 \email{gpb@lanl.gov}
\affiliation{Theoretical Division, Los Alamos National Laboratory,
Los Alamos, NM 87544, USA }

\date{\today}

\begin{abstract}

We study the dynamics of dephasing in a quantum two-level system by
modeling both $1/f$ and high-frequency noise by random telegraph
processes. Our approach is based on a so-called spin-fluctuator model
in which a noisy environment is modeled by a large number of
fluctuators. In the continuous limit we obtain an effective random
process (ERP) that is described by a distribution function of the
fluctuators. In a simplified model, we reduce the ERP to the two (slow and fast) ensembles
of fluctuators. Using this model, we study
decoherence in a superconducting flux qubit and we compare our
theoretical results with the available experimental data. We
demonstrate good agreement of our theoretical predictions with the
experiments. Our approach can be applied to many quantum systems, such as biological complexes,
semiconductors, superconducting and spin qubits, where the  effects of interaction with the
environment are essential.

\end{abstract}



\preprint{LA-UR-12-10060}

\maketitle


\maketitle

\section{ Introduction}

Decoherence is one of the main obstacles for
building useful quantum devices. Understanding the mechanisms of
decoherence and achieving long decoherence times is crucial for many fields
of science and applications including quantum computation and quantum information
\cite{nielsen},
protein dynamics \cite{FFMP,FC,YF}, dynamics of excitons and charge separation in biological
complexes \cite{ECR,SMR,MSL,CWW,CBM}, and the new
and rapidly growing fields of NMR and MRI with ultra small (microtesla)
magnetic fields \cite{EMV,ZMV,CHMM}. In the latter case, the Larmor
frequencies of the spin precession become relatively small (in the kHz
region), causing the effects of $1/f$ noise become so important that noise suppression must be
used.

In many situations the influence of noise can be modeled by an
ensemble of two-level systems or fluctuators
\cite{BGA,GABS,MSAM,SSMM,FAMP,ICJM,YBLA}. Depending on the distribution
of parameters of the fluctuators, such as amplitudes and switching rates, and
coupling constants, this model can describe both
Gaussian and non-Gaussian effects of noise \cite{BGA,GABS,MSAM}.
Recent experiments with Josephson qubits \cite{CSMM,SLHN,APNY}, on the quantum dynamics of
excitons in light-harvesting antennas in photosynthetic complexes \cite{ECR,SMR,MSL,CWW,CBM}
demonstrated these important contributions of noise and thermal fluctuations to decoherence,
relaxation processes, and quantum coherence effects.

In this paper, we study relaxation and dephasing processes using a spin-fluctuator model
\cite{BGA,GABS}. In the spin-fluctuator model, fluctuations are described by a random telegraph
process (RTP) produced by $N$ fluctuators. Each
fluctuator is characterized by two parameters: its amplitude and
switching rate. Depending on the distribution function of
fluctuators over amplitudes and switching rates, the RTP can describe
noise for a broad range of frequencies using  spectral characteristics that include both low-
and
high-frequencies noise.

We consider the noisy environment produced by a large number of fluctuators, $N \gg 1$. In the
limit $N\rightarrow \infty$, we obtain an effective random process (ERP) described by a
continuous distribution of fluctuators. We
derive a closed system of integro-differential equations for functions averaged over the ERP.
Even though this system of equations is closed, it is still very complicated for direct analysis
and even for numerical solutions.

We study two approximations in which these equations are reduced to a system of differential
equations. The first one we call the Gaussian approximation, because, as we demonstrate, in the
simplest case of a two-level system (qubit) under the influence of an ERP, it yields the
relation:  $\langle \exp
(i\varphi)\rangle=\exp(-{\langle\varphi^2/2\rangle})$, where
$\varphi$ is the random angle of the Bloch vector. This Gaussian
approximation is widely used in theoretical and experimental
research to descre the influence of noise on quantum systems
\cite{MSAM,ICJM,YHNN,BMAH,HJBJ,KRDD,MNAL}. In many situations, this approximation is very useful
because (i) it captures some important properties of noisy dynamics and (ii) it is simple to
apply. However, the Gaussian approximation does not describe different ``non-Gaussian" effects
which can play a significant role.

Our second approximation  is based on two effective
fluctuators which include both low and high frequency noisy components. We show
that this approximation goes beyond the Gaussian approach and
better describes the  experimental results for a superconducting flux
qubit in a noisy environment \cite{YHNN}.

\subsection*{\bf Our main results}

\begin{itemize}
    \item We create a new model based on an effective random process (ERP) that includes
        both
        slow
    (low-frequencies) and fast (high-frequencies) fluctuators. This model can describe
     the influence of noise  on a quantum system over a wide frequency range.

    \item For the functions averaged over the ERP, we obtain an
    integro-differential  master equation which we reduce to a closed system of differential
    equations in two approximations: (i) a Gaussian approximation and (ii) an approximation
    of two-effective fluctuators. Both of these approximations describe, to some extent, the
    contributions from low and high frequency noise.

    \item We demonstrate that the two-effective fluctuator approximation accurately models
        ``non-Gaussian" effects observed in experiments with  superconducting flux qubits
        \cite{YHNN} .

    \item We show that the two-effective-fluctuators model better
    describes the suppression of $1/f$ noise in experiments involving echo decay in
    superconducting flux qubits \cite{YHNN}.
\end{itemize}

This paper is organized as follows.  In Sec. II, a general model of
noise based on ERP is introduced that describes both low-frequency ($1/f$) and high-frequency
noise. In Sec. III, we use the reduced density matrix approach to describe the interaction of a
quantum system with its environment by a master equation. For two cases (i) the Gaussian
approximation and (ii) the two effective- fluctuator model, we reduce the system of
integro-differential equations to a closed system of
differential equations.  In Sec. IV, the general method developed in
Secs. II and III is applied to describe the decoherence of a
superconducting flux qubit for free induction decay and for echo decay
experiments. In the same Sec. IV, we compare our theoretical predictions with available
experimental data and demonstrate a good agreement with experiments. We
conclude in Sec. V with a discussion of our results. In the Appendices we present some technical
details.

\section{Description of noise using a random telegraph process}

To describe noise we use the spin-fluctuator model developed in
\cite{BGA,GABS}. In this model, noise is described by a sum of $N$ uncorrelated fluctuators,
$\xi_N= \sum^N_{i=1}\zeta_i(t)$, where  $\zeta_i(t)$ is a random telegraph process (RTP). The
variable, $
\zeta_i(t)$, takes the values,  $-a_i$ or $a_i$. Consequently, $\zeta^2_i(t)= a^2_i = \rm
const$.

The RTP obeys following relations \cite{BGA,KV1,KV2,KV3}
\begin{eqnarray}\label{C1a}
\langle  {\zeta}_i(t) \rangle = 0 , \\
\langle  {\zeta}_i(t) {\zeta}_j(t') \rangle = \delta_{ij}a^2_i
e^{-2\gamma_i|t-t'|}. \label{C1}
\end{eqnarray}
The amplitude, $a_i$,  together with the switching rate, $\gamma_i$,
completely characterize the $i$-th fluctuator. The correlation function related to $\xi_N(t)$ is
defined as,
$\chi_N(|t-t'|)= \langle \xi_N(t)\xi_N(t')\rangle $.
Using Eqs. (\ref{C1a}) and (\ref{C1}), we obtain
\begin{eqnarray}
\chi_N(|t-t'|) =  \sum^{N}_{i=1}a^2_i e^{-2\gamma_i|t-t'|}.
\end{eqnarray}
Further, assuming  $N\gg 1$, we consider continuous
distributions of amplitudes and switching rates. The corresponding
correlation function,  $\chi(t)=\lim_{N\rightarrow \infty}
\chi_N(t)$, can be written as
\begin{eqnarray}
\chi(|t-t'|)  = \iint dw(\sigma,\gamma)  \sigma^2   e^{-2\gamma|t-t'|},
\label{CF5r}
\end{eqnarray}
where, $\sigma^2= \lim_{N\rightarrow \infty }Na^2$, and  $dw(\sigma,\gamma)$  depends on the
specific distributions of amplitudes and switching rates. The random process described by the
function,
$\xi(t)=\lim_{N\rightarrow \infty} \xi_N(t)$, we call an effective random process (ERP).

In order to model the characteristic behavior of the spectral density of noise in different
frequency domains,
we introduce a family of random variables and distributions, $\{\xi_n(t),dw_n(\sigma,\gamma)\}$.
In particular,  $n=1$ corresponds to low-frequency ($1/f$) noise, and $n=2$ corresponds to the
Lorentzian spectrum for high frequencies. (See Appendix C for details.)

Accordingly, we introduce the ERP as $\xi(t) = \sum_n \xi_n(t)$, where each $\xi_n(t)$ is an
independent source of noise. This implies $\langle \xi_m(t)\xi_n(t')\rangle =0$ $(m\neq n)$.
As shown in the Appendix A, the corresponding spectral density $S_n(\omega)$ behaves as
$S_n(\omega) \sim 1/\omega^n$ in some region of frequencies. The total correlation function is a
sum of the partial correlation functions, $\chi(|t-t'|)=\sum_n\chi_{n}(|t-t'|)$, where
\begin{align}
  \chi_n(|t-t'|) =  \iint dw_n(\sigma,\gamma)  \sigma^2   e^{-2\gamma|t-t'|}.
 \label{ccn}
\end{align}

In this paper, we adopt the simple model introduced in \cite{BGA} for uncorrelated $\sigma$ and
$\gamma$. We define the distribution function as
\begin{align}
dw_n(\sigma,\gamma)  =\delta(\sigma-  \sigma_n){\mathcal P}_n(\gamma) d\sigma d\gamma,
\end{align}
where $\sigma_n$ is a typical value of the amplitude and
\begin{align}\label{Neq11}
&{\mathcal P}_n(\gamma)d\gamma=A_n\Theta(\gamma_{c_n} - \gamma)\Theta
(\gamma - \gamma_{m_n})\displaystyle\frac{d\gamma}{\gamma^n}.
 \end{align}
Here, $\Theta(x)$, is a step-function; and  $\gamma_{m_n}$ and $\gamma_{c_n}$ are the lower and
upper switching rates, respectively. The normalization constant, $A_n$, is:
\begin{align}
A_n =\left \{  \begin{array}{ll}
\displaystyle \frac{1}{\ln(\gamma_{c_1}/\gamma_{m_1})}, & n=1\\
&\\
\displaystyle \frac{(n-1)\gamma^{n-1}_{m_n}}{(1- \gamma^{n-1}_{m_n}/
\gamma^{n-1}_{c_n})}, & n \neq 1
\end{array}
\right .
\end{align}

In the following, we restrict ourselves to two
important cases: $n=1$ and $n=2$, which are related to $1/f$ noise
and to high-frequency noise with the corresponding spectral densities. (The case for arbitrary
$n$ is analyzed in Appendix C.) We denote $\gamma_m =\gamma_{m _1}$, $\gamma_c =\gamma_{c _1} $,
and $\gamma_0 =\gamma_{c _2} $.  For the distribution functions, ${\mathcal P}_1(\gamma)$ and
${\mathcal
P}_2(\gamma)$,  we impose  conditions at the point
$\gamma = \gamma_c$, so that $\gamma_{m _2}= \gamma_{c _1} =
\gamma_c$ $(\gamma_m <\gamma_c <\gamma_0)$.

Using  Eq. (\ref{Chi_1}) (Appendix C), we obtain,
\begin{eqnarray}
\chi_{1}(\tau) = \sigma^2_{1} A_1(E_1(2\gamma_m \tau) - E_1(2\gamma_c \tau) ), \\
\chi_{2}(\tau) = \sigma^2_{2} A_2 \bigg (\frac{ E_2(2\gamma_c \tau)}{\gamma_c} - \frac{
E_2(2\gamma_0 \tau)}{\gamma_0} \bigg).
\label{CF3}
\end{eqnarray}
Computation of the spectral density,
\begin{eqnarray}
S_{n}(\omega) = \frac{1}{\pi}\int\limits_{0}^{\infty}\chi_{n}(\tau) \cos(\omega \tau) d\tau ,
\label{Sf1c}
\end{eqnarray}
yields
\begin{align}\label{SF1}
&S_{1}(\omega) = \frac{\sigma^2_{1} A_1}{\pi\omega}\bigg( \arctan
\Big(\frac{\omega}{2\gamma_m}\Big)- \arctan\Big(\frac{\omega}{2\gamma_c}\Big)\bigg), \\
&S_{2}(\omega) = \frac{\sigma^2_{2} A_2}{\pi\omega^2}\ln\bigg(\frac{1 + \omega^2/4\gamma_c^2}{1
+
\omega^2/4\gamma_0^2}\bigg),
\label{SF2}
\end{align}
where $A_1= 1/\ln(\gamma_c/\gamma_m)$ and $A_2= \gamma_c/(1- \gamma_c/\gamma_0)$.
\begin{figure}[tbh]
\begin{center}
\scalebox{0.35}{\includegraphics{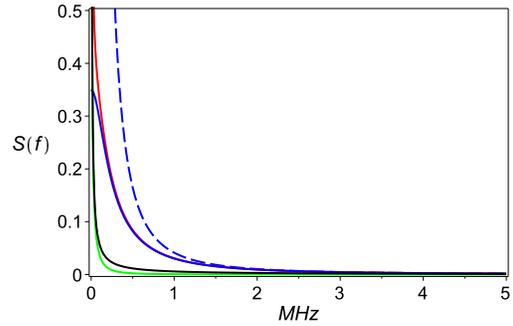}}
\end{center}
\caption{Spectral density of noise ($2\gamma_m = 1\,\rm s^{-1} $, $2\gamma_c
= 1\,\rm \mu s ^{-1}$, $2\gamma_0 = 10\,\rm \mu s ^{-1}$, $\sigma_1 = \sigma_2 = 1$, $\omega =
2\pi f$). Red line: The total spectral density, $S(f)=S_1(f)+ S_2(f) $. Green line: Contribution
of the slow fluctuators described by $S_1(f)$. Blue line: Contribution of the fast fluctuators
described by $S_2(f)$. Blue dashed line: The Lorentzian spectrum, $S_{L}(f)\sim 1/f^2$. Black
line: The spectral density of $1/f$ noise, $S_{1/f}(f)= A/(2\pi f)$. }
 \label{S_1}
\end{figure}

From Eqs. (\ref{SF1}) and (\ref{SF2}) it follows that, in the interval, $\gamma_m <  \omega
<\gamma_c$, the spectral density $S_{1}(\omega)$  describes  $1/f$ noise. Indeed, in this
interval $S_{1}(\omega) \approx A/\omega$, where $A= \sigma^2_{1}/(2\ln(\gamma_c/\gamma_m)) $,
and $\gamma_m$ and $\gamma_c$  are related to the infrared, $\omega_m= 2\gamma_m$, and the
ultraviolet, $\omega_c = 2\gamma_c$, frequency cutoffs, respectively (Appendix C). For
$S_2(\omega)$, we obtain the following asymptotic behavior: $S_2(\omega) \sim 1/\omega^2$
($\omega \gg \omega_c$). Thus, asymptotically $S_2(\omega)$ has a Lorentzian spectrum.

To estimate the relative contributions of different processes for low- and
high-frequency noise, we evaluate the relation,
$S_{2}(\omega)/S_{1}(\omega)$ at frequencies $\omega
\approx 0$ and   $\omega \approx \omega_c$. A simple computation yields
the following rough estimate:
\begin{eqnarray}
\frac {S_{2}(\omega)}{S_{1}(\omega)}\sim \left \{\begin{array}{ll}
\displaystyle\frac{\sigma^2_{2}  }{ \sigma^2_{1}}\frac{ \gamma_m  \ln(\gamma_c/\gamma_m)}{
\gamma_c},&   \omega \approx 0,\\
\\
\displaystyle \frac{\sigma^2_{2}  }{ \sigma^2_{1}}  \ln(\gamma_c/\gamma_m), &  \omega \approx
\omega_c .
\end{array}
\right .
\end{eqnarray}
Taking  values typical for superconducting qubits, $2\gamma_m
\approx 1 \,\rm s^{-1}$ and $2\gamma_c \approx 1\, {\rm \mu s}^{-1}$, we obtain
\begin{eqnarray}
\frac {S_{2}(\omega)}{S_{1}(\omega)}\sim \left \{\begin{array}{ll}
\displaystyle 10^{-5}{\sigma^2_{2}  }/{ \sigma^2_{1}},&   \omega \approx 0,\\
\\
\displaystyle 10\,{\sigma^2_{2}  }/{ \sigma^2_{1}}, &  \omega \approx \omega_c .
\end{array}
\right .
\end{eqnarray}
Thus,  for low-frequency noise the main contribution near $\omega
=0$ is provided by slow fluctuators (SF) with the switching rates, $\gamma$, being in the
interval $(\gamma_m,\gamma_c)$.
However, for high frequencies, $\omega \gtrsim \omega_c $, the
contribution of fast fluctuators (FF) with $\gamma \gtrsim \gamma_c$
dominates. (See Fig. \ref{S_1}.)

\section{Master equation for averaged density matrix}

We consider a quantum system governed by the Hamiltonian, ${\mathcal H(t)}$ (generally
time-dependent), depending on control
parameters, $\lambda_i$ ( external flux, biased current, critical current, etc). The noise
associated with  fluctuations of these parameters is described by
random functions,  $\delta\lambda_i(t)$. For simplicity, we restrict ourselves to only one
fluctuating parameter, $\delta\lambda(t)$, denoting it as $\xi(t)$. Generalization for many
parameters is straightforward. Expanding  the Hamiltonian to first order in $\xi(t)$, we obtain
\begin{eqnarray} \label{Neq1}
{\mathcal H(t)} = {\mathcal H}_0(t) +{\mathcal V}(t)\xi(t).
\end{eqnarray}

To include the effects of a thermal bath, we use the reduced density matrix approach leading to
the master equation:
\begin{eqnarray}\label{Meq1}
\frac{d\rho}{dt} =- i[{\mathcal H(t)}, \rho] +{\mathcal L}\rho,
\end{eqnarray}
where  the superoperator, ${\mathcal L}$, describes coupling to the bath.

Using (\ref{Neq1}), one can recast Eq. (\ref{Meq1}) as
\begin{eqnarray}\label{Meq2a}
\frac{d\rho(t)}{dt} =- i[{\mathcal H}_0(t), \rho(t)] +{\mathcal L}\rho(t) \nonumber \\
- i [\xi(t)  {\mathcal V}(t), \rho(t)].
\end{eqnarray}
For the average density matrix, this yields
\begin{eqnarray}\label{Meq1b}
\frac{d\langle\rho(t)\rangle}{dt} =-i [{\mathcal H}_0(t),
 \langle\rho(t)\rangle] +{\mathcal L}\langle\rho(t)\rangle \nonumber \\
 -i[{{\mathcal V}}(t), \langle  X(t)\rangle] ,
 \label{Ideq}
\end{eqnarray}
where $\langle  X(t)\rangle= \langle { \xi}(t) \rho(t)\rangle $, and the average $\langle
{\;}\rangle$ is taken over the random process describing the noise.

As before, we assume that fluctuations are produced by
the ERP, so that  $\xi(t) =\sum_n\xi_n(t)$, and the correlation function can be written as a sum
of the partial correlation functions, $\chi(|t-t'|)=\sum_n\chi_{n}(|t-t'|)$. (See Eq.
\ref{ccn}.)

Eq. (\ref{Meq1b}), an integro-differential equation, is rather complicated. However, in two
important cases (the Gaussian
approximation and the approximation by effective fluctuators) we
obtain a closed system of first order differential equations
(Appendix B). Below we summarize our results for $\xi(t)=\xi_1(t)+ \xi_2(t)$, where $\xi_1(t)$
is
related to slow fluctuators leading to $1/f$ noise, and $\xi_2(t)$ is related to fast
fluctuators
leading to high-frequency noise.\\

{\em The Gaussian approximation.} Applying the method described in Appendix B, we find that, in
the Gaussian approximation, the master equation can be recast as follows
\begin{align}\label{Meq1br}
\frac{d\langle \rho(t)\rangle}{dt} =& -i [{\mathcal H}_0(t),
 \langle\rho(t)\rangle] +{\mathcal L}\langle\rho(t)\rangle \nonumber\\
&-[{\mathcal V }(t),[ {\mathcal K}(t) , \langle   \rho(t)\rangle]] + { \mathcal O}
(\parallel{\mathcal V}\parallel^4) ,
\end{align}
where ${\mathcal K}(t) = \int_0^t dt'\chi(t-t') U^\dagger(t) \tilde{\mathcal V}(t')  U(t)$. We
denote $\tilde {\mathcal V}(t)= U(t)  {\mathcal V}(t) U^\dagger(t)$, and
\begin{align}
 U(t) = T\Big(e^{i\int\limits_{0}^{t}{\mathcal H}_0 (t')dt'}\Big).
\end{align}

{\em The approximation by two effective fluctuators.} In the approximation by two effective
fluctuators, the set of slow, $\xi_1(t)$, and fast, $\xi_2(t)$, fluctuators is approximated by
two effective fluctuators: one for SF and the other for FF. The total correlation function,
$\chi(|t-t'|)   = \chi_{1}(|t-t'|) + \chi_{2}(|t-t'|)$ , is approximated as
\begin{align}
 \chi_{n}(|t-t'|)\approx {a^\ast_n}^2
e^{-2\gamma^\ast_n|t-t'|} , \quad n=1,2 ,
\label{C6}
\end{align}
where $a_n^\ast$  and $\gamma^\ast_n$ (the effective amplitude and switching rate) are defined
as
follows: ${a^\ast_n}^2 = \chi_n(0)$ and $\gamma^\ast_n = -(1/2)d\ln \chi(t)/dt|_{t=0}$. (For
details, see Appendix B.)

Applying the method developed in Appendix B for an arbitrary  system
of stochastic first-order ordinary differential equations, we
obtain from Eq. (\ref{Meq1b}) the following closed system of ordinary
differential equations:
\begin{widetext}
\begin{eqnarray}
\frac{d}{dt}{\langle\rho(t)\rangle}&=&-i [{\mathcal H}_0(t), {\langle\rho\rangle}] +{\mathcal
L}{\langle\rho\rangle} - i [{\mathcal V}(t), \langle X_1(t)\rangle] - i[{\mathcal V}(t) ,\langle
X_2(t)  \rangle] , \\
\frac{d}{dt}{\langle  X_{ 1}(t)\rangle}&=&- 2\gamma^\ast_{ 1}\,{ \langle X_{1}(t)\rangle} -i
[{\mathcal H}_0(t), {\langle X_{ 1}(t)\rangle}] +{\mathcal L}{\langle X_{ 1}(t)\rangle}
\nonumber
\\
&&-i{a_1^\ast}^2 [{{\mathcal V}} (t), \langle\rho(t)\rangle] -i[{{\mathcal V}} (t),\langle  X_{
12}(t)\rangle ],\\
\frac{d}{dt}{\langle  X_{ 2}(t)\rangle}&=&- 2\gamma^\ast_{ 2}\,{ \langle X_{2}(t)\rangle} -i
[{\mathcal H}_0(t), {\langle X_{ 1}(t)\rangle}] +{\mathcal L}{\langle X_{ 2}(t)\rangle}
\nonumber
\\
&&-i{ a_2^\ast}^2 [{{\mathcal V}} (t), \langle\rho(t)\rangle] -i[{{\mathcal V}} (t),\langle  X_{
12}(t)\rangle ], \\
\frac{d}{dt}{\langle  X_{ 12}(t)\rangle}&=&- 2(\gamma^\ast_{ 1} +\gamma^\ast_{ 2}){ \langle
X_{12}(t)\rangle} -i [{\mathcal H}_0(t), {\langle X_{ 12}(t)\rangle}] +{\mathcal L}{\langle X_{
12}(t)\rangle} \nonumber \\
&&-i {a_1^\ast}^2 [{{\mathcal V}} (t),\langle  X_{ 2}(t)\rangle -i{a_2^\ast}^2[{{\mathcal V}}
(t),\langle  X_{ 1}(t)\rangle ],
\end{eqnarray}
\end{widetext}
where $\langle {  X}_1 (t) \rangle = \langle \xi_1(t) {  \rho} (t)
\rangle $,
 $\langle {  X}_2 (t) \rangle = \langle \xi_2(t) {  \rho} (t) \rangle $ and
 $\langle {  X}_{12} (t) \rangle = \langle \xi_1(t)  \xi_2(t) {
\rho} (t) \rangle $.

\section{Non-Gaussian noise and decoherence in a superconducting phase qubit}

In this section, the general method developed in Secs. 2 and 3  is applied
to describe relaxation effects in a superconducting qubit. The
effective Hamiltonian for a superconducting qubit can be written as
\cite{YHNN} (see also references therein),
\begin{eqnarray}\label{Hqb}
{\mathcal H}(t)= -\frac{1}{2}{\boldsymbol {\Omega(t) }}\cdot
\boldsymbol \sigma,
\end{eqnarray}
where $\boldsymbol \sigma$ denotes the Pauli matrices. We assume that
${\mathcal H(t)}$ depends on the control parameters, $\lambda_i$, of
the system, including external flux, biased current, critical current, etc.
Limiting ourselves to a single fluctuating parameter, $\lambda$, and
expanding the Hamiltonian in Eq. (\ref{Hqb}) to first order in the
fluctuations, $ \delta \lambda(t)$, we obtain,
\begin{eqnarray}\label{Hqb1}
{\mathcal H}(t)=-\frac{1}{2}{\boldsymbol {\Omega }}\cdot \boldsymbol
\sigma -  \frac{1}{2} \delta \lambda(t)\frac{\partial
 \boldsymbol {\Omega }}{\partial \lambda}\cdot \boldsymbol \sigma,
\end{eqnarray}
where, for simplicity, we assume that ${\boldsymbol {\Omega }}$
does not depend on $t$. In the eigenbasis of the unperturbed  Hamiltonian, Eq. (29) takes the
form
\begin{align}\label{Hqb2}
{\mathcal H}(t)=&-\frac{1 }{2} \Omega\sigma_z -  \frac{ 1 }
{2}D_{\lambda,z} \delta \lambda(t) \sigma_z \nonumber \\
& - \frac{1 }{2} D_{\lambda,\bot} \delta \lambda(t) \sigma_\bot,
\end{align}
where $D_{\lambda,z} = \partial \Omega/\partial \lambda$  and
$\sigma_\bot$  denotes the transverse spin components, either
$\sigma_x$ or $\sigma_y$. (We adopt the notation of
Ref. \cite{ICJM}.)

Below, in the framework of the ERP model, we obtain the relaxation rates, and compare our
results
with the results which follow from the well-known Bloch-Redfield (BR) theory
\cite{BF1,Rag} applied to the external noise \cite{ICJM}. Before proceeding,  we present here
some important results of the BR approach.

In BR theory, the dynamics of a two-level system is described by two rates: the longitudinal
relaxation rate, $\Gamma_1 = T_1^{-1}$, and the
transverse relaxation rate, $\Gamma_2 = T_2^{-1}$. BR theory is
valid if $T_1,T_2 \gg \tau_c$, where $\tau_c$ is the fluctuation correlation
time. The transverse relaxation rate, $\Gamma_2$, is
a combination of $\Gamma_1$ and the so-called ``pure dephasing''
rate, $\Gamma_\varphi$,
\begin{eqnarray}
\Gamma_2 = \frac{1}{2}\Gamma_1 + \Gamma_\varphi.
\end{eqnarray}
In terms of the  spectral density of noise,
$S_\lambda(\omega)$, these rates are defined as follows \cite{ICJM}:
\begin{eqnarray}
\label{BR8}
\Gamma_1 =  \pi D^2_{\lambda,\bot} S_\lambda(\Omega), \\
\Gamma_\varphi =  \pi D^2_{\lambda,z} S_\lambda(0). \label{BR3}
\end{eqnarray}

In our approach, fluctuations of the parameter, $\lambda$, are
described by an ERP. Thus, $\delta \lambda(t) = \sum_n\xi_n(t)$. Further, we restrict ourselves
to consideration only the case, $n=1,2$. Then, $\delta \lambda(t) = \xi_1(t) + \xi_2(t)$, where
$\xi_1(t)$ describes the contribution to the ERP of SF, and $\xi_2(t)$ describes the
contribution
of FF. The spectral density of noise can be written as $S_\lambda(\omega) =S_1(\omega)
+S_2(\omega)$.

Since only FFs have small correlation times and satisfy the  conditions
of applicability of BR theory, we use the spectral density of FFs  given by Eq. (\ref{SF2})  to
calculate the relaxation and dephasing rates provided by the BR theory. We obtain
\begin{widetext}
\begin{eqnarray}
\Gamma_1 =  \pi D^2_{\lambda,\bot} S_{2}
(\Omega) =  D^2_{\lambda,\bot}\frac{\sigma^2_{2} \gamma_c}
{\Omega^2 (1-{\gamma_c}/{\gamma_0})}\ln\bigg(\frac{1 +
\Omega^2/4\gamma_c^2}{1 + \Omega^2/4\gamma_0^2}\bigg), \\
 \Gamma_\varphi = \pi D^2_{\lambda,z}S_{2}(0) = D^2_{\lambda,z}
 \frac{\sigma^2_{2} }{4 \gamma_c}\bigg(1+\frac{\gamma_c}{\gamma_0}\bigg).
\end{eqnarray}
\end{widetext}
Note, that the validity of the BR theory is restricted by the
condition:
 $\Gamma_1 \tau_{2},\Gamma_2 \tau_{2} \ll 1$,
where $\tau_{2}   = ({1}/{2\gamma_c}) (1+ {\gamma_c}/{\gamma_0})$
is the  effective correlation time of the FF.

The above effective rates can also be obtained directly from the
averaged expressions for the partial rates,
\begin{eqnarray} \label{BR2}
\Gamma_1 =D^2_{\lambda,\bot}\int\limits_{\gamma_c}^{\gamma_0}\frac{
{\sigma}_2^2\gamma}{4\gamma^2
+ \Omega^2}\,dw_2(\gamma), \\
 \Gamma_\varphi =D^2_{\lambda,z}\int\limits_{\gamma_c}^{\gamma_0}\frac{{\sigma}_2^2
 }{2\gamma}\,dw_2(\gamma).
 \label{BR2a}
\end{eqnarray}

\subsection{Pure decoherence}

Let us consider the Hamiltonian (\ref{Hqb2}) for a pure decoherence
case. Then $D_{\lambda,\bot}=0$, and the Hamiltonian ${\mathcal H}(t)$ takes the form
\begin{eqnarray} \label{EqH2}
{\mathcal H}(t) =-\frac{1}{2}\Omega(t)\sigma_z  - \frac{1}{2}
D_{\lambda,z}\delta \lambda(t) \sigma_z,
\end{eqnarray}
where $D_{\lambda,z}=  \partial \Omega/\partial \lambda $.

The equation of motion  for  the density matrix, $i\dot\rho=
[\mathcal H(t),\rho]$, reduces to only one component:
\begin{align}
\frac{d}{dt}{\rho}_{01} = i\Omega(t) {\rho}_{01} + i D_{\lambda,z}\delta \lambda(t){\rho}_{01}.
\label{MD1}
\end{align}
The matrix elements, ${\rho}_{00}$ and  ${\rho}_{11}$, are constant.

Setting ${\rho_{01}(t)} = \tilde \rho_{01}(t)e^{i\varphi_0(t)}$, where $\varphi_0(t)=\int^t_0
\Omega dt $ is a regular phase, we find that $ \tilde \rho_{01}(t)$ satisfies the following
differential equation:
\begin{eqnarray}
\frac{d}{dt}{\tilde \rho}_{01}(t) =i  D_{\lambda,z}\delta \lambda(t) \tilde \rho_{01}(t).
\label{F1}
\end{eqnarray}
Its  solution can be written as
\begin{eqnarray}
{\tilde \rho}_{01}(t) = e^{i\varphi(t)}\rho_{01}(0),
\end{eqnarray}
where $\varphi(t) = \int\limits_{0}^{t} D_{\lambda,z}\delta \lambda(t)dt$ is a random phase.
After averaging over the random process, we
obtain $\langle{\tilde \rho}_{01}(t)\rangle = \langle e^{i\varphi(t)}\rangle\rho_{01}(0)$. This
yields
$\langle{\rho}_{01}(t)\rangle =  e^{i\varphi_0(t)} \langle e^{i\varphi(t)}\rangle\rho_{01}(0)$.
Thus, the problem of obtaining exact solution of Eq. (\ref{MD1}) reduces to the computation of
the generating functional, $\langle e^{i\varphi(t)}\rangle$.

Returning to Eq. (\ref{MD1}), one can see that, for averaged components of the density matrix,
it
takes the form
\begin{align}
\frac{d}{dt}\langle{\rho}_{01}(t)\rangle =
i\Omega(t)\langle{\rho}_{01}(t)\rangle
+ i D_{\lambda,z}\langle \delta \lambda(t){\rho}_{01}(t)\rangle.
\label{MD1a}
\end{align}

In what follows, we obtain  solutions of Eq. (\ref{MD1a}) in the Gaussian approximation  and in
the two-effective-fluctuator approximation. We apply these solutions to describe two widely used
experimental protocols: (i) free induction decay and
(ii) echo decay. We compare our theoretical predictions with the
experimental data \cite{YHNN}, and demonstrate that the experimental
results (i) are described by the Gaussian approximation and (ii) that
the details of the dynamics of the signal decay can be understood by
using slow and fast effective fluctuators. We also demonstrate that
the  approach based on two effective fluctuators allows one to fit
the experimental data better.

\subsubsection{Gaussian approximation}

In the Gaussian approximation Eq. (\ref{MD1a}) can be presented in the form
\begin{eqnarray}
\frac{d}{dt}\langle{\rho}_{01}(t)\rangle =
i\Omega(t)\langle{\rho}_{01}(t)\rangle \nonumber \\
 - D^2_{\lambda,z}\Big(\int\limits_{0}^{t}\chi(t -t')  d t'\Big)
\langle{\rho}_{01}(t)\rangle,
\end{eqnarray}
where $\chi(t -t') =\langle \delta \lambda(t) \lambda(t')\rangle$. Its solution can be
written as
\begin{eqnarray}
\langle {\rho}_{01} (t) \rangle=  e^{i\varphi_0(t)}  e^{-(1/2)\langle
\varphi^2(t)\rangle} \langle {\rho}_{01}(0) \rangle,
\end{eqnarray}
where $\varphi_0(t) = \int_0^t \Omega(\tau) d \tau$ is a regular phase, $\varphi(t) =
D_{\lambda,z} \int\limits_{0}^{t}\delta\lambda(t')dt'$
is the random phase accumulated during the time, $t$, and
\begin{eqnarray}
 \langle \varphi^2(t) \rangle = D^2_{\lambda,z} \int\limits_{0}^{t}
 \int\limits_{0}^{t}\chi(|t' -t''|) dt'd t'' ,
 \label{varphi}
\end{eqnarray}
is the variance of $\varphi(t)$.

Thus, in the Gaussian approximation,
the random phase of the free-induction decay is Gaussian distributed, and we obtain the
well-known result for the generating functional,
\begin{eqnarray}
\langle e^{i\varphi(t)}\rangle = e^{-(1/2)\langle \varphi^2(t)\rangle}
\label{FID5}.
\end{eqnarray}
Using the spectral function of noise, $S_\lambda(\omega)$, one can rewrite (\ref{FID5}) as
\cite{BGA,ICJM}
\begin{align}
\langle e^{i\varphi(t)}\rangle = \exp
\Big ( -\frac{t^2}{2} D^2_{\lambda,z}\int\limits_{-\infty}^{\infty} d\omega S_\lambda (\omega)
{\rm sinc}^2 \frac{\omega t}{2}\Big ),
\label{FID5b}
\end{align}
where ${\rm sinc}\,x = \sin x/x$.

In the echo experiments, the total  phase, $\psi(t)$, is defined as difference between two free
evolutions, so that \cite{BGA,ICJM}
\begin{align}
\psi(t) =  D_{\lambda,z}\Bigg(\int\limits_{0}^{t/2}\delta\lambda(t')dt' -
\int\limits_{t/2}^{t}\delta\lambda(t')dt'\Bigg).
\label{Echo1}
\end{align}
In the Gaussian approximation, we obtain
\begin{align}
\langle e^{i\psi(t)}\rangle = e^{-(1/2)\langle \psi^2(t)\rangle},
\label{ES1a}
\end{align}
where
\begin{align}
\langle \psi^2(t)\rangle =  D^2_{\lambda,z} \int\limits_{0}^{t}\int\limits_{0}^{t}dt'
d t'' \chi(|t' -t''|) \nonumber \\
- 4D^2_{\lambda,z} \int\limits_{0}^{t/2} dt' \int\limits_{t/2}^{t} d t'' \chi(|t'
-t''|) . \label{varpsi}
\end{align}

In terms of the spectral density, the echo decay can be written as
\cite{BGA,ICJM}
\begin{widetext}
\begin{eqnarray}
\langle e^{i\psi(t)}\rangle = \exp
\Big ( -\frac{t^2}{2} D^2_{\lambda,z}\int\limits_{-\infty}^{\infty} d\omega S_\lambda (\omega)
\sin^2\frac{\omega t}{4}
{\rm sinc}^2\frac{\omega t}{4}\Big ).
\label{ES2}
\end{eqnarray}
\end{widetext}
In Appendix C, we obtain explicit expressions for $\langle
\varphi^2(t)\rangle$ and $\langle \psi^2(t)\rangle$.

Using the asymptotic formulae for the exponential integrals,  $E_n(z)$,
\cite{abr}, we find that, for $\gamma_m t \ll 1$ ($\gamma_c t < 1$), the
free-induction decay produced by SF is given by
\begin{align}
\langle e^{i\varphi(t)}\rangle = \exp
\bigg ( -{t^2} D^2_{\lambda,z} A\Big(\ln \frac{1}{2\gamma_m t} +
{\mathcal O}(1)\Big)\bigg ),
\label{FID5d}
\end{align}
where  $A= \sigma^2_{\xi_1}/(2\ln(\gamma_c/\gamma_m))$. Substituting $\omega_m = 2 \gamma_m$, we
find that (\ref{FID5d}) is exactly the same expression that is used in the literature for
estimating the quasistatic contribution of $1/f$ noise to the free-induction decay \cite{ICJM}.
In the same limit, for the echo decay we obtain
\begin{align}
\langle e^{i\psi(t)}\rangle = \exp  ( -t^2 D^2_{\lambda,z} A\ln 2 ),
\end{align}
which coincides with the corresponding formula obtained from Eq.
(\ref{ES2}) for $\omega_m t\ll1$ \cite{CA}.

The contribution of low frequencies, $\omega t \ll 1$,  in  (\ref{FID5}) obtained in the limit
$\gamma_mt , \gamma_c t\ll1$, is
\begin{eqnarray}
\langle e^{i\varphi(t)}\rangle = \exp
\bigg ( -\frac{t^2}{2} D^2_{\lambda,z} \sigma_{\xi_1}^2\bigg ).
\label{FID5e}
\end{eqnarray}
This coincides with the corresponding expression widely used in the literature \cite{ICJM}.

\subsubsection{Two-effective fluctuator model}

In this Section, the SF and FF introduced above are approximately
described by two effective fluctuators with the following
correlation functions (see Appendix B),
\begin{eqnarray}
 \chi_n(|t-t'|)  \approx \sigma_n^2 e^{-2\gamma_n |t - t'|},
   \quad (n=1,2).
  \label{ccc}
\end{eqnarray}
where
\begin{eqnarray}
\gamma_n=-\frac{1}{2} \frac{\partial\ln( \chi_{n}(\tau))}{\partial
 \tau}\bigg|_{\tau=0},
\end{eqnarray}
the effective amplitude and  switching rate being $\sigma_n $ and $\gamma_n$. Computation yields
\begin{align}
{\gamma}_{1}= \frac{\gamma_c- \gamma_m}{\ln(\gamma_c/\gamma_m)}  , \\
{\gamma}_{2}= \frac{\gamma_c\ln(\gamma_0/\gamma_c)}{1-\gamma_c/\gamma_0}.
\end{align}

For the averaged functions: $ \langle { \rho_{01}} (t) \rangle$,
$\langle { X}_1 (t) \rangle = \langle \xi_1(t) { \rho_{01}} (t)
\rangle $, $\langle { X}_2 (t) \rangle = \langle \xi_2(t) {
\rho_{01}} (t) \rangle $ and  $\langle { X}_{12} (t) \rangle =
\langle \xi_1(t) \xi_2(t) { \rho_{01}} (t) \rangle $, we obtain
the following closed system of first-order differential equations:
\begin{widetext}
\begin{align}\label{X1}
\frac{d}{dt} \langle { \rho_{01}} (t) \rangle = i\Omega(t)\langle { \rho_{01}} (t) \rangle
+  i  D_{\lambda,z}(\langle { X}_1 (t) \rangle + \langle { X}_2 (t) \rangle), \\
\frac{d}{dt}\langle {X}_1(t) \rangle = -2\gamma_1\langle {X}_1(t) \rangle + i\Omega(t) \langle {
X}_1 (t) \rangle
 + i  D_{\lambda,z} (\langle { X}_{12}(t) \rangle +a^2_1\langle { \rho_{01}}(t) \rangle),  \\
 \frac{d}{dt}\langle {X}_2(t) \rangle = -2\gamma_2\langle {X}_1(t) \rangle + i\Omega(t) \langle
 {
 X}_2 (t) \rangle
 + i  D_{\lambda,z} (\langle { X}_{12}(t) \rangle +a^2_2\langle { \rho_{01}}(t) \rangle),  \\
 \frac{d}{dt}\langle {X}_{12}(t) \rangle = -2(\gamma_1+\gamma_2)\langle {X}_{12}(t) \rangle +i
 \Omega(t) \langle { X}_{12} (t) \rangle
 + i  D_{\lambda,z} (a^2_2\langle { X}_{1}(t) \rangle + a^2_1\langle { X}_2(t) \rangle).
\label{X1a}
\end{align}
\end{widetext}

The solution of Eqs. (\ref{X1}) - (\ref{X1a}) can be written as,
\begin{eqnarray} \label{SL1}
\langle   \rho_{01}(t)\rangle = e^{i\varphi_0(t)}\Phi_1(t) \Phi_2(t)
\rho_{01}(0),\\
i  D_{\lambda,z}\langle  {X}_{1}(t) \rangle =e^{i\varphi_0(t)} {\dot \Phi}_1(t) \Phi_2(t)
\rho_{01}(0),\\
i  D_{\lambda,z}\langle  {X}_{2}(t) \rangle =e^{i\varphi_0(t)} \Phi_1(t){\dot \Phi}_2(t)
\rho_{01}(0),\\
  D^2_{\lambda,z}\langle  {X}_{12}(t) \rangle = - e^{i\varphi_0(t)}{\dot \Phi}_1(t){\dot
  \Phi}_2(t)
\rho_{01}(0),
\end{eqnarray}
where $\varphi_0(t)= \int^t_0 \Omega(t') dt' $. We denote by $\Phi_i(t) $ $(i=1,2)$ the
generating functional  of the RTP \cite{KV1,KV2,KV3},
\begin{eqnarray}
\Phi_i(t) = \Big\langle \exp\Big \{i\int\limits_{0}^{t}
d\tau \xi_i(\tau) v_i\Big \}\Big\rangle,
\end{eqnarray}
where $v_i^2 =  D^2_{\lambda,z} a_i^2$. The generating functional  satisfies the second order
differential equation \cite{KV1,KV2,KV3},
\begin{equation}\label{SF4}
  \frac{d^2\Phi_i }{dt^2}+ 2\gamma_i  \frac{d\Phi_i }{dt}+{ v_i^2 } \Phi_i=0,
\end{equation}
with the initial conditions being $\Phi_i(0)=1$ and $d\Phi_i(0)/dt=0$.

{\em Free induction and echo decay solutions for a single fluctuator.}
In the following, we consider solutions of Eq. (\ref{SF4})
corresponding to free induction signal and echo signal experiments. Previously  Eq. (\ref{SF4})
was studied in \cite{BGA,GABS1,GABS2}.
\begin{itemize}
\item The solution corresponding to the decay of the free
induction signal is given by \cite{BGA,PFFF,GABS2}
\begin{eqnarray}\label{Req4}
\Phi^f_i(t)= \frac{e^{-\gamma_i t}}{\mu_i}\sinh({\gamma_i \mu_i t}) \nonumber \\
+e^{-\gamma_i t}\cosh({\gamma_i \mu_i t}),
\label{FD_1}
\end{eqnarray}
where $\mu_i = \sqrt{1-{ v_i}^2/\gamma_i^2}$.
\item  In the echo experiments,  the $\pi$-pulse with duration,
$\tau_1$, is applied at time, $\tau =t/2$, to switch the two states
of qubit. It is assumed that $\tau_1\ll\tau$. The corresponding
solution for the functional $\Phi^e_i(t)$,
with the initial conditions $ \Phi^e_i(0) =1$ and $d\Phi^e_i(0) /dt=0$, is written as
\cite{BGA}
\begin{align}
\Phi^e_i(t) =&\frac{e^{-\gamma_i t}}
{\mu_i^2}
\big(\mu_i \sinh({\gamma_i \mu_i t}) \nonumber \\
&+ \cosh({\gamma_i \mu_i t})+ \mu_i^2 -1 \big). \label{SE1}
\end{align}
\end{itemize}

\subsection{Comparison with experiment}

In this section, we compare our theoretical predictions with the
experimental data obtained in \cite{YHNN} and the theoretical
results of the model \cite{ZJR}.  The measurement of the decoherence due
to $1/f$ noise   was done for the flux qubit described by the
effective Hamiltonian \cite{YHNN},
\begin{eqnarray}
H_0 =- \frac{\varepsilon}{2}\sigma_z - \frac{\Delta}{2}\sigma_x,
\end{eqnarray}
with the energy difference between two eigenstates  $E_{01}=
 \sqrt{\varepsilon^2 + \Delta^2}$.

The diagonalized Hamiltonian, with the fluctuations only of $E_{01}$, can be written as
\begin{eqnarray}
H =- \frac{E_{01}}{2}\sigma_z  -
\frac{1}{2}\sum_a D_{\lambda_a,z}\delta\lambda_a(t) \,\sigma_z ,
\end{eqnarray}
where $ D_{\lambda_a,z}={\partial E_{01}}/{\partial\lambda_a }$, and
the term, $\delta\lambda_a(t)$, describes the fluctuations of $\lambda_a$ in the Hamiltonian.
In
the experiments \cite{YHNN}, the authors  studied the decoherence due to
fluctuations of (i) the normalized external flux, $n_\phi =
\Phi_{\rm ex}/\Phi_0$, where $\Phi_0$ is the flux quantum, and (ii)
the SQUID bias current, $I_b$.  The contributions from different
decoherence sources were separated, so that the fluctuations of
$n_\phi$ and $I_b$ were observed independently.

We consider two approximations: (i) the two-effective  fluctuator
approximation and  (ii) the Gaussian approximation.

\subsubsection{Two-effective-fluctuator approximation}

In the approximation of two effective fluctuators, $\delta\lambda (t)=\xi_1(t)  + \xi_2(t) $,
and
\begin{eqnarray}
\langle \xi_i(t)  \xi_j(t') \rangle  =\delta_{ij} \sigma^2_i
e^{-2\gamma_i t} , \quad i=1,2.
\end{eqnarray}

Our  task is to determine the fitting parameters: $( v_1,\gamma_1,
v_2,\gamma_2$), where $v_i^2 =   D^2_{\lambda,z}\sigma_i^2$, and
\begin{align}
{\gamma}_{1}= \frac{\gamma_c(1- \gamma_m/\gamma_c)}{\ln(\gamma_c/\gamma_m)}  , \quad
{\gamma}_{2}= \frac{\gamma_c\ln(\gamma_0/\gamma_c)}{1-\gamma_c/\gamma_0}.
\end{align}
The switching rates, $\gamma_m$ and $\gamma_c$, are chosen according to the available
experimental data for the spectral behavior of $1/f$ noise, namely, $\gamma_m \sim 1\;\rm
s^{-1}$
and $\gamma_c \sim 1\;\rm \mu s^{-1}$. Then, the only two free fitting parameters are $\gamma_0$
and $v_2$. Their values are chosen from the best fit of theoretical results to the experimental
data.

To fix the value of $v_1$, we use the experimental data
for echo decay fitted to the Gaussian decay,
$\exp({-\big(\Gamma^g_{\varphi E}(\lambda) t)^2}\big)$, using the relation
from \cite{YHNN,CA}
\begin{eqnarray}
\Gamma^g_{\varphi E}(\lambda) = \sqrt{A_{\lambda }\ln 2}
\bigg|\frac{\partial E_{01}}{\partial\lambda }\bigg|. \label{G1b}
\end{eqnarray}
The constant, $A_{\lambda}$, is determined from the experimental data describing the behavior of
the spectral density of $1/f$ noise, $S_{\lambda } (\omega) =A_{\lambda }/\omega $, at the
frequency, $f = 1 \rm Hz$ \cite{YHNN}.

Inserting $A_{\lambda}=  \sigma_1^2/(2\ln (\gamma_c/\gamma_m))$ into Eq. (\ref{G1b}), we obtain
\begin{eqnarray}
{v_1(\lambda)} = \Gamma^g_{\varphi E} (\lambda)\sqrt{\frac{2
 \ln(\gamma_c/\gamma_m)}{\ln2}}.
\label{G2}
\end{eqnarray}
\begin{figure}[tbh]
\begin{center}
\scalebox{0.35}{\includegraphics{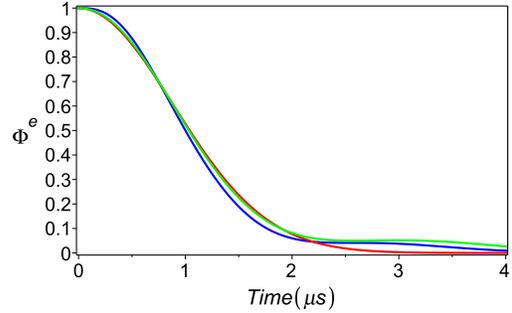}}
\end{center}
\caption{Sample A from \cite{YHNN}. Echo decay,  $\Phi^e(t) = \langle e^{i\psi(t)}
\rangle$. The blue solid line is fit by two-fluctuator solution, $\langle e^{i\psi(t)} \rangle =
\Phi_1^e(t)\Phi_2^e(t)$. The green solid  line is the theoretical predictions of Ref.
\cite{ZJR}.
Red solid line corresponds to the Gaussian decay, $\langle e^{i\psi(t)} \rangle=
e^{-(\Gamma^g_{\varphi E} t)^2}$. The experimental data (not shown) are obtained for
decoherence
at value  $\Delta n_\phi = 0.0009$ (Fig. 4a, Ref. \cite{YHNN}). }
\label{EXP1}
\end{figure}
In  Fig. \ref{EXP1}, we compare our theoretical predictions with the
experimental data obtained for decoherence of a flux qubit with
fluctuations of the external normalized flux, $n_\phi$ (sample A from
\cite{YHNN}). To fit our theoretical results to the experimental
curves, we set two cutoffs for ${1/f}$ noise as: $\gamma_c = 0.5 \;\rm \mu s^{-1}$ and
$\gamma_m=
0.5 \;\rm s^{-1}$. Then, calculating the  switching
rate, $\gamma_1$ , we obtain $\gamma_1 = 0.04 \;\rm \mu s^{-1}$.  From Fig.
4c (in \cite{YHNN}), describing echo dephasing rate
 $\Gamma^g_{\varphi E}$ vs $\Delta n_\phi$, we extract $\Gamma^g_{\varphi E} =  0.8 \;\rm \mu
 s^{-1}$, and, then, using $(\ref{G2})$, we obtain $ v_1 =
\sqrt{v^2_{n_\phi}} =4.92 \;\rm \mu s^{-1}$. The parameters, $v_2$ and
$\gamma_0$, are chosen by best fitting our curve to the experimental
data.  For the high-frequency noise we obtain the upper cutoff as
$\gamma_0= 4.25 \;\rm \mu s^{-1}$. For the switching rate $\gamma_2 $ this
yields:  $\gamma_2=1.2 \;\rm \mu s^{-1}$. The amplitude $v_2$ we choose as
$v_2 =2.72 \;\rm \mu s^{-1}$.
\begin{figure}[tbh]
\begin{center}
\scalebox{0.35}{\includegraphics{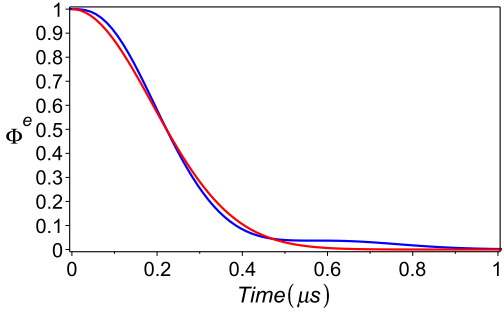}}
\end{center}
\caption{Sample B from \cite{YHNN}. Echo decay,  $\Phi^e(t) = \langle e^{i\psi(t)} \rangle$. The
blue solid line is fit by the two fluctuator solution. The red solid line corresponds to the
Gaussian decay $ e^{-(\Gamma^g_{\varphi E} t)^2}$. The experimental data are obtained for
decoherence for value $\Delta n_\phi =-0.0008$ (Fig. 4d, Ref. \cite{YHNN}). }
\label{EXP1b}
\end{figure}
In  Fig. \ref{EXP1b}, we compare our theoretical
predictions with the experimental data obtained for decoherence in a
flux qubit with fluctuations of the external flux, $n_\phi$ (sample B from
\cite{YHNN}). The fitting parameters obtained in the same way as for sample A  are: $ \gamma_1 =
0.04 \;\rm \mu s^{-1} $ ,  $  v_1 = 21 \;\rm \mu s^{-1} $, $ \gamma_2 = 5.75 \;\rm \mu s^{-1} $,
$  v_2 = 12.45 \;\rm \mu s^{-1} $ and $\Gamma^g_{\varphi E}= 3.75 \;\rm \mu s^{-1}$. In Figs.
\ref{EXP1} and \ref{EXP1b}, the echo decay of  the two fluctuator model (blue curves) resulted
from both low and high frequency
fluctuators.
\begin{figure}[tbh]
\begin{center}
\scalebox{0.35}{\includegraphics{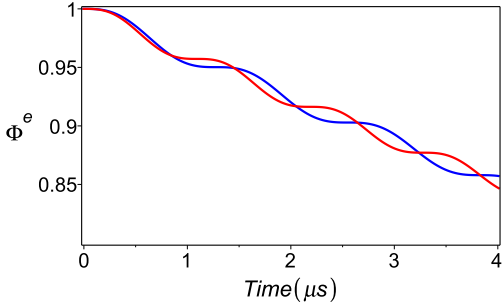}}
\end{center}
\caption{ Suppression of  $1/f$ noise in echo-decay experiment
in a flux qubit at fluctuations of the external flux. The blue solid line corresponds to sample
A, and the red solid line corresponds to
sample B from Ref. \cite{YHNN}.} \label{EXP1c}
\end{figure}

In order to determine the contribution from only $1/f$ noise to this echo decay,
we present in Fig. \ref{EXP1c} the decay (for samples A and B) provided by only a slow effective
fluctuator  with the same parameters,
$\gamma_1$ and $v_1$,  as those indicated in Figs. \ref{EXP1} and \ref{EXP1b}. One can see from
Fig. \ref{EXP1c} that in both cases, a suppression of $1/f$
noise is up to $95 \%$ in the time-interval, $(0 - 1\mu s)$.
\begin{figure}[tbh]
\begin{center}
\scalebox{0.35}{\includegraphics{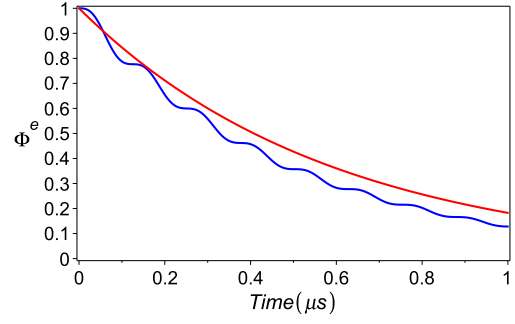}}
\end{center}
\caption{ Echo decay,  $\Phi^e(t) = \langle e^{i\psi(t)} \rangle$. The blue
solid line is fit by the solution of the two-effective-fluctuators
model, with the choice of $ \gamma_1 = 0.04\rm MHz $, $   v_1 = 10.5 \rm MHz $, $ \gamma_2 = 2
\rm MHz $ and $   v_2 = 50 \rm MHz $. The red solid line corresponds to the exponential decay,
$e^{-\Gamma_{\varphi E} t}$ with $\Gamma_{\varphi E}= 1.7 \rm MHz $. The experimental data are
obtained for  decoherence at value of SQUID bias current $I_b = -0.7 \mu \rm A$ (sample A, Fig.
3c,  Ref. \cite{YHNN}). \label{EXP2}}
\end{figure}

In Fig. \ref{EXP2}, we compare our theoretical results for echo
decay with the experimental data obtained for decoherence in a flux
qubit for fluctuations of SQUID bias currents $I_b$ (sample A from
\cite{YHNN}). In all considered cases, we find that our solutions based on two
effective fluctuators better fit  the experimental data than the
theoretical description of the Gaussian approximation
used in \cite{YHNN}.
\begin{figure}[tbh]
\begin{center}
\scalebox{0.325}{\includegraphics{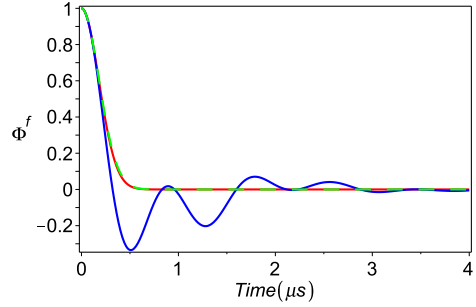}}
\end{center}
\caption{Sample A from \cite{YHNN}. Free induction signal  decay, $\Phi^f(t) = \langle
e^{i\varphi(t)} \rangle$. Blue solid line fits  to the
two fluctuators solution, $\Phi^f(t) =\Phi_1^f(t)\Phi_2^f(t)$. Red solid line corresponds to the
Gaussian decay, $\Phi^f= e^{-(\Gamma^g_{\varphi F} t)^2}$ with $\Gamma^g_{\varphi F}= 3.97 \;\rm
\mu s^{-1} $. Green dashed line presents the Gaussian approximation for free decay, $\langle
e^{i\varphi(t)}\rangle = e^{-(1/2)\langle \varphi^2(t)\rangle} $. (Data are taken from Ref.
\cite{YHNN} for  decoherence at various flux biases $n_\phi$, sample A.) \label{FID1}}
\end{figure}

We consider also free induction decay, and compare the obtained
effective decoherence rate, $\Gamma^g_{\varphi F}$, with the
experimental data and theoretical results of the Gaussian
model \cite{ICJM,YHNN}. For free induction decay, the two-fluctuator
solution is
\begin{eqnarray}
\langle e^{i\varphi(t)} \rangle=\Phi^f_1(t)\Phi^f_2(t),
\label{FD_2}
\end{eqnarray}
where $\Phi^f_i(t)$ $(i=1,2)$ are given by Eq. (\ref{FD_1}).

Expanding (\ref{FD_2}) into a Taylor series, we obtain
\begin{eqnarray}
\langle e^{i\varphi(t)} \rangle = 1 -\frac{1}{2}(v_1^2 + v_2^2) t^2 + \dots .
\end{eqnarray}
Then, comparing with the Gaussian decay, $e^{-(\Gamma^g_{\varphi F} t)^2}$, we obtain
\begin{eqnarray}
\Gamma^g_{\varphi F}= \sqrt{\frac{1}{2}(v_1^2 + v_2^2)}.
\label{G1}
\end{eqnarray}
Substituting $v_1=4.92 \rm MHz$ and $v_2=2.72 \rm MHz$
(sample A), we find $\Gamma^g_{\varphi F}= 3.97 \rm MHz$.

Computation for the sample A of the the ratio,  $\Gamma^g_{\varphi
F}/\Gamma^g_{\varphi E}$,  yields $\Gamma^g_{\varphi
F}/\Gamma^g_{\varphi E}\approx 4.96$. This is in a
good agreement with the theoretical prediction, $\Gamma^g_{\varphi
F}/\Gamma^g_{\varphi E}\lesssim  5$ \cite{ICJM}, and with the
estimate from the experimental data yielding a ratio between
$4.5$ and $7.5$ \cite{YHNN}.

\subsubsection{Gaussian approximation}

In the Gaussian approximation, free induction signal decay is described by
\begin{align}
\langle e^{i\varphi(t)}\rangle = e^{-(1/2)\langle \varphi^2(t)\rangle},
\end{align}
where $\langle {\varphi^2(t)} \rangle =\langle {\varphi_1^2(t)}
\rangle +\langle {\varphi_2^2(t)} \rangle $, and $ \langle \varphi_{n}^2(t) \rangle$ ($n=1,2$)
is given by Eq. (\ref{C7}).

In Fig. \ref{FID1}, we compare the Gaussian approximation (green dashed line), the
two-fluctuator
solution (blue solid line) and the Gaussian decay (red solid line). As one can see, the Gaussian
approximation and the Gaussian decay yield practically the same results. However, the
two-fluctuator solution shows non-Gaussian oscillatory behavior.

We also considered the echo decay signal for the Sample A. Our theoretical results for
echo decay follow from (\ref{ES1a}) and (\ref{varpsi})
\begin{eqnarray}
 \langle e^{i\psi(t)} \rangle =
e^{-(1/2)\langle {\psi^2(t)} \rangle },
\end{eqnarray}
where $\langle {\psi^2(t)} \rangle =\langle {\psi_1^2(t)}
\rangle +\langle {\psi_2^2(t)} \rangle $, and $ \langle \psi_{{n}}^2(t) \rangle$ (n=1,2) is
given
by Eq. (\ref{C8})
\begin{figure}[tbh]
\begin{center}
\scalebox{0.35}{\includegraphics{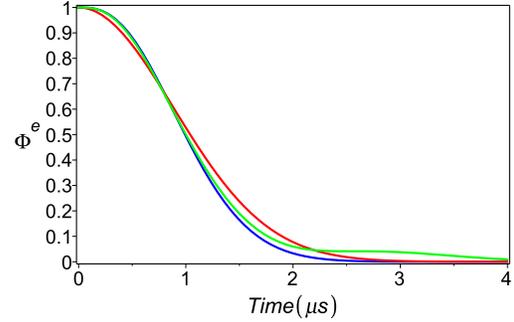}}
\end{center}
\caption{Sample A \cite{YHNN}. Echo signal decay, $\Phi^e(t) = \langle e^{i\psi(t)}\rangle$. The
blue solid line is fit by the Gaussian approximation, $\langle e^{i\psi(t)} \rangle =
e^{-(1/2)\langle {\psi^2(t)} \rangle }$. Green solid line: two fluctuator solution, $\langle
e^{i\psi(t)}\rangle = \Phi_1^e(t)\Phi_2^e(t)$. The red solid line corresponds to Gaussian decay,
$\langle e^{i\psi(t)} \rangle= e^{-(\Gamma^g_{\varphi E} t)^2}$.} \label{Gauss_1}
\end{figure}

In Fig. \ref{Gauss_1}, we present theoretical results (sample A) for echo
decay for: the two-effective-fluctuator model (green curve); the Gaussian
approximation (blue curve), and Gaussian decay used in [19] (red
curve).

\begin{figure}[tbh]
\begin{center}
\scalebox{0.35}{\includegraphics{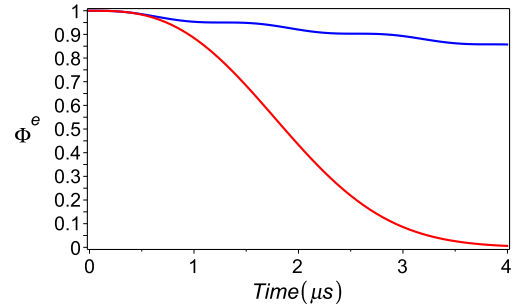}}
\end{center}
\caption{Suppression of  $1/f$ noise in echo-decay experiment. The blue
solid line corresponds to the two fluctuator solution with only slow fluctuators. The red solid
line corresponds to the Gaussian approximation with slow fluctuators.
}
\label{Gauss_3}
\end{figure}

In Fig. \ref{Gauss_3}, we compare the theoretical results for suppression of
$1/f$ noise by slow fluctuators in the two-effective-fluctuator model and
the Gaussian approximation. One can see that, up to $1 \; \rm \mu s$, both
descriptions give similar results. For times larger the 1 microsecond, the
Gaussian approximation does not describe the suppression of $1/f$
noise, since fluctuators with $\gamma \approx \gamma_c$ dominate.

\section{Conclusions}

The approach based on modeling of noisy environment by an ensemble of two-level systems
(fluctuators) is widely used for quantum solid-state systems
\cite{BGA,GABS,MSAM,SSMM,ICJM,YBLA}. Recent experiments with Josephson phase qubits
\cite{CSMM,SLHN,APNY} demonstrated the importance of noise of all frequencies in  decoherence
processes and stimulated theoretical discussions on the contributions of low- and
high-frequencies fluctuators \cite{ICJM}.

In this paper, we have discussed the SF model for continuous distribution of fluctuators to
describe both, low- and high-frequency noise. We considered two approximations of our model: the
Gaussian approximation and two-fluctuator approximation, and  compared our theoretical
predictions with the experimental results for decoherence of a superconducting flux qubit
\cite{YHNN}. We showed a good agreement between our theoretical model and experimental results.

We should emphasize that the two-fluctuator approximation leads to the non-Gaussian behavior in
the signal decay. The non-Gaussian effects, yielding contribution to a particular behavior of the
tail in the spin echo signal, are very strong for free induction signal decay. The main problem
is that available experimental results on superconducting qubits (including those reported in
Ref. \cite{YHNN}) may not have a good enough precision to distinguish Gaussian and non-Gaussian
behavior. However, it is no doubt that the non-Gaussian behavior is relevant to many situations
and can help to understand better the nature of noise and its action on the system under
consideration.

\section*{Acknowledgments}

This work was carried out under the auspices of the
National Nuclear Security Administration of the U.S. Department of
Energy at Los Alamos National Laboratory under Contract No.
DE-AC52-06NA25396. This research  was partly supported by the Intelligence
Advanced Research Projects Activity (IARPA). All statements of fact,
opinion or conclusions contained herein are those of the authors and
should not be construed as representing the official views or
policies of IARPA, the ODNI, or the U.S. Government. ~  A.I. Nesterov
acknowledges the support from the CONACyT, Grant No. 118930, and IARPA and the Quantum Institute
through the CNLS at LANL.

\begin{widetext}
\appendix

\section{Some properties of random processes }

\subsection{Random telegraph process}

In this section, we derive some useful formulae for the random telegraph
process (RTP) defined by
 $\xi_N(t) = \sum^N_{i=1}\zeta_i(t)$ with the correlation function,
$\chi_{N}(|t-t'|)= \langle \xi_N(t)\xi_N(t')\rangle $, given by
\begin{eqnarray}
\chi_{N}(|t-t'|) =  \sum^{N}_{i=1}a^2_i e^{-2\gamma_i|t-t'|}.
\end{eqnarray}
We assume that the RTP is described by $N$ uncorrelated fluctuators,
$  \zeta_i(t)$. Each fluctuator switches randomly between the values
$-1$ and $1$ with the probability $1/2$, so that $\zeta^2_i(t) = a^2_i = \rm const$, and after
averaging over the initial states of each fluctuator, the following correlation relations hold
\cite{KV1,KV2,KV3}
\begin{align}\label{A1}
&\langle  {\zeta}_i(t) \rangle =0 , \\
&\langle  {\zeta}_i(t) {\zeta}_j(t') \rangle =  \delta_{ij}a^2_i
e^{-2\gamma_i(t-t')},  \;  t \geq t',\\
&M^{i}_n (t_1,t_2,\dots, t_n) = a^2_i e^{-2\gamma_i|t_1-t_2|}M^{i}_{n-2} (t_3,\dots, t_n),
\label{A1a}
\end{align}
where
\begin{align}
M^{i}_n (t_1,t_2,\dots, t_n) = \langle  {\zeta}_i(t_1)\dots {\zeta}_i(t_n)
\rangle, \quad
 t_1\geq t_2\geq \dots \geq t_n.
\end{align}

From Eqs. (\ref{A1}) -  (\ref{A1a}), a recursive formula follows
\begin{align}\label{A2}
M^{N}_n (t_1,t_2,\dots, t_n) = \chi_{N}(t_1-t_2) M^N_{n-2} (t_3,\dots, t_n),
\end{align}
where
\begin{eqnarray}
M^N_n (t_1,t_2,\dots, t_n) = \langle  {\xi_N}(t_1)\dots \xi_{N}(t_n) \rangle, \quad
 t_1\geq t_2\geq \dots \geq t_n.
\label{A2a}
\end{eqnarray}

The RTP is conveniently described by the generating functional \cite{KV3},
\begin{eqnarray}
\Phi_N[t;v(\tau)] = \Big\langle \exp\Big \{i\int\limits_{0}^{t}
 d\tau \xi_N(\tau) v(\tau)\Big \}\Big\rangle .
\label{A3}
\end{eqnarray}
Applying Eq. (\ref{A2})  and using the Taylor expansion  of Eq.
(\ref{A3}), we obtain an exact integral equation for the generating
functional $\Phi_N[t;v]$:
\begin{eqnarray}
\Phi_N[t;v(\tau)]  = 1 - \int\limits_{0}^{t} dt_1
\int\limits_{0}^{t_1} dt_2 \chi_{N}(t_1-t_2) v(t_1) v(t_2)
\Phi_N[t_2;v(\tau)].
\end{eqnarray}
One can transform this integral equation into the integro-differential equation,
\begin{align}
\label{A4} \frac{d}{dt}\Phi_N[t;v(\tau)] = -
v(t)\int\limits_{0}^{t} dt_1  \chi_{N}(t-t_1) v(t_1)
\Phi_N[t_1;v(\tau)]
\end{align}

Let $R[t;\xi_N(\tau)]$ be an arbitrary functional. Then, using Eq.
(\ref{A2}) and a Taylor expansion in $\xi_N(\tau)$, one can show
that the following  correlation splitting formula holds:
\begin{eqnarray}
\langle  {\xi_N}(t_1){\xi_N}(t_2)R[t;\xi_N(\tau)] \rangle =
\chi_{N}(t_1-t_2)\langle R[t;\xi_N(\tau)] \rangle, \;  t_1\geq
t_2\geq \tau. \label{A4r}
\end{eqnarray}
To calculate the correlator $\langle  {\xi_N}(t)R[t;\xi_N(\tau)] \rangle $ for $\tau\leq t$ we
use the following relations \cite{KV3}:
\begin{eqnarray}
 \langle  {\xi_N}(t)R[t;\xi_N(\tau) +\eta(\tau)] \rangle = \Big\langle{\xi_N}(t) \exp\Big
 \{\int\limits_{0}^{t} d\tau \xi_N(\tau)
 \frac{\delta}{\delta \eta(\tau)}\Big \}\Big\rangle R[t;\eta(\tau)],
\end{eqnarray}
where $\eta(\tau)$ is a deterministic function. With the help of Eq. (\ref{A4}), we obtain
\begin{align}
 \langle  {\xi_N}(t)R[t;\xi_N(\tau) +\eta(\tau)] \rangle
 =&\int\limits_{0}^{t} dt_1 \chi_N(t-t_1) \Big\langle\frac{\delta}{\delta \eta(t_1)} \exp\Big
 \{\int\limits_{0}^{t_1} d\tau \xi_N(\tau
  \frac{\delta}{\delta \eta(\tau)}\Big \}\Big\rangle R[t;\eta(\tau)] \nonumber \\
 =&\int\limits_{0}^{t} dt_1 \chi_N(t-t_1) \Big\langle\frac{\delta}{\delta \eta(t_1)}
     R[t; \eta(\tau)+\xi_N(\tau)\Theta(t_1 - \tau)]\Big \rangle .
\end{align}
Taking the limit $\eta \rightarrow 0$, we find
\begin{align}
 \langle  {\xi_N}(t)R[t;\xi_N(\tau) ] \rangle
 =\int\limits_{0}^{t} dt_1 \chi_N(t-t_1) \Big \langle \frac{\delta}{\delta \xi_N(\tau)} \tilde{
 R}[t,t_1; \xi_N(\tau)] \Big \rangle ,
\end{align}
where
\begin{align}\label{A5}
 \tilde{ R}[t,t_1; \xi_N(\tau)] = R[t; \xi_N(\tau)\Theta(t_1 - \tau +0)] .
\end{align}
By differentiating  (\ref{A5}) with respect to time $t$, we obtain
\begin{align}
\frac{d}{dt}\langle  {\xi_N}(t)R[t;\xi_N(\tau) ] \rangle  -
 \int\limits_{0}^{t} dt_1  \frac{d}{dt}\chi_N(t-t_1) \frac{\delta}{\delta \xi_N(\tau)}
 \Big \langle \tilde{ R}[t,t_1; \xi_N(\tau)]\Big  \rangle
=\Big\langle  {\xi_N}(t)\frac{d}{dt}R[t;\xi_N(\tau) ] \Big\rangle .
\label{A6a}
\end{align}
This formula  generalizes the differential formula \cite{KV1,KV2,KV3}
\begin{align}
\Big(\frac{d}{dt} +2\gamma \Big)\langle  {\zeta}(t)R[t;\zeta(\tau) ]
\rangle =\Big\langle  {\zeta}(t)\frac{d}{dt}R[t;\zeta(\tau)]
\Big\rangle ,
\end{align}
taking place for the RTP described by $\zeta(t)$ with switching
rate, $\gamma$.

{\bf Theorem 1.}  For the random telegraph process, ${\xi_N}(t)$, the following relation holds:
\begin{align}
\langle  {\xi_N}(t')R[t;\xi_N(\tau) ] \rangle = \frac{\chi_N(t'-t)}{\chi_N(0)} \langle
{\xi_N}(t)R[t;\xi_N(\tau) ] \rangle,
\quad t' \geq t, \label{A20}
\end{align}
where $R[t;\xi_N(\tau) ] $ is an arbitrary functional.
\bigskip

{\em Proof.} Writing $\xi^2_N(t)$ as
\begin{eqnarray}
\xi^2_N(t) = \sum\limits_{i=1}^{N} \zeta^2_i(t) + \sum\limits_{i\neq
j}^{N} \zeta^2_i(t)\zeta^2_j(t),
\end{eqnarray}
we can employ the fact that $\zeta^2_i(t) = \rm const$
\cite{KV1,KV2}. Next, using the relation $\chi_N(0)
=\sum\limits_{i=1}^{N} \xi^2_i(t) $, we obtain
\begin{eqnarray}
\frac{\xi^2_N(t)}{\chi_N(0)} -\frac{1}{\chi_N(0)}\sum\limits_{i\neq
j}^{N} \zeta_i(t) \zeta_j(t) =1. \label{A21}
\end{eqnarray}
Inserting $(\ref{A21})$ into the l.h.s. of Eq. (\ref{A20}), we find
\begin{align}
\langle  {\xi_N}(t')R[t;\xi_N(\tau) ] \rangle =  \langle {\xi_N}(t')
\frac{\xi^2_N(t)}{\chi_N(0)} R[t;\xi_N(\tau) ] \rangle
- \frac{1}{\chi_N(0)}\sum\limits_{i\neq j}^{N} \langle  \zeta_i(t)
\zeta_j(t){\xi_N}(t')R[t;\xi_N(\tau) ] \rangle .
\end{align}
Then, applying (\ref{A2}), we obtain
\begin{align}
\langle  {\xi_N}(t')R[t;\xi_N(\tau) ] \rangle = \frac{\chi_N(t'-t)}{\chi_N(0)} \langle
\xi_N(t) R[t;\xi_N(\tau) ] \rangle
- \frac{1}{\chi_N(0)}\sum\limits_{i\neq j}^{N}\langle  \zeta_i(t)
\zeta_j(t) \rangle \langle {\xi_N}(t')R[t;\xi_N(\tau) ] \rangle ,
\quad t' \geq t.
\end{align}
Since for, $i\neq j$, $\langle \zeta_i(t) \zeta_j(t) \rangle =0$,
this yields
\begin{align}
\langle  {\xi_N}(t')R[t;\xi_N(\tau) ] \rangle = \frac{\chi_N(t'-t)}{\chi_N(0)}
\langle \xi_N(t) R[t;\xi_N(\tau) ], \quad
 t' \geq t.
\end{align}
{\em Corollary.} In the limit $N\rightarrow \infty$, one has
\begin{align}
\langle  {\xi}(t')R[t;\xi(\tau) ] \rangle = \frac{\chi(t'-t)}{\chi(0)}
 \langle \xi(t) R[t;\xi(\tau) ], \quad
 t' \geq t.
\end{align}
where $\chi(t'-t)=\lim_{N\rightarrow \infty}\chi_N(t'-t)$.

\subsection{Effective Random Process}

We define the effective random telegraph process (ERP) for  $N\gg 1$, as
$\xi(t)=\lim_{N\rightarrow \infty} \xi_N(t)$, considering the continuous
distribution of amplitudes and switching rates. The
correlation function,  $\chi(t)=\lim_{N\rightarrow \infty}
\chi_N(t)$, can be written as
\begin{eqnarray}
\chi(|t-t'|)  =  \lim_{N\rightarrow \infty} \sum^{N}_{i=1}a^2_i e^{-2\gamma_i|t-t'|}=\iint
dw(\sigma,\gamma)  \sigma^2   e^{-2\gamma|t-t'|},
\label{CF5}
\end{eqnarray}
where, $\sigma^2= \lim_{N\rightarrow \infty }Na^2$, and  $dw(\sigma,\gamma)$,  depends on the
specific distribution functions of fluctuators on the amplitudes and switching rates. The main
relations for the ERP can be obtained from the previous section by taking the limit $N\rightarrow
\infty$. Below we present the most important formulae.

The generating functional for the ERP being defined as
\begin{eqnarray}
\Phi[t;v(\tau)] = \Big\langle \exp\Big \{i\int\limits_{0}^{t}
 d\tau \xi(\tau) v(\tau)\Big \}\Big\rangle
\label{A3r}
\end{eqnarray}
satisfies the following integral equation:
\begin{eqnarray}
\Phi[t;v(\tau)]  = 1 - \int\limits_{0}^{t} dt_1
\int\limits_{0}^{t_1} dt_2 \chi(t_1-t_2) v(t_1) v(t_2)
\Phi[t_2;v(\tau)].
\label{A5r}
\end{eqnarray}
One can transform  Eq. (\ref{A5r}) into the integro-differential equation,
\begin{align}
\label{A4ra}
\frac{d}{dt}\Phi[t;v(\tau)] = -
v(t)\int\limits_{0}^{t} dt_1  \chi(t-t_1) v(t_1)
\Phi[t_1;v(\tau)]
\end{align}

For an arbitrary functional $R[t;\xi(\tau)]$ the following correlation splitting formula holds:
\begin{eqnarray}
\langle  {\xi}(t_1){\xi}(t_2)R[t;\xi(\tau)] \rangle =
\chi(t_1-t_2)\langle R[t;\xi(\tau)] \rangle, \;  t_1\geq
t_2\geq \tau. \label{A6r}
\end{eqnarray}
Finally, the differentiation formula (\ref{A6a}) takes the form
\begin{align}
\frac{d}{dt}\langle  {\xi}(t)R[t;\xi(\tau) ] \rangle  -
 \int\limits_{0}^{t} dt_1   \frac{d}{dt}\chi(t-t_1) \frac{\delta}{\delta \xi(\tau)}
  \Big \langle\tilde{ R}[t,t_1; \xi(\tau)]\Big  \rangle
=\Big\langle  {\xi}(t)\frac{d}{dt}R[t;\xi(\tau) ] \Big\rangle .
\end{align}
{\em Relation to the Gaussian random process.} We would like to mention here an important
consequence of the central limit
theorem concerning a relation between ERP and the Gaussian random
process. Assume that for individual fluctuators the correlation
relations are given by
\begin{eqnarray}
\langle  {\zeta}_i(t) \rangle = 0 , \\
\langle  {\zeta}_i(t) {\zeta}_j(t') \rangle = \frac{\sigma^2}{N}\delta_{ij}
e^{-2\gamma|t-t'|}.
\label{C5}
\end{eqnarray}
Then, for $N\rightarrow \infty$, the ERP, defined by $ \xi_N(t)$,
becomes a Gaussian Markovian process with an exponential correlation function \cite{KV1,KV3}
\begin{eqnarray}
\langle  {\xi}(t) {\xi}(t') \rangle = \sigma^2e^{-2\gamma|t-t'|},
\label{C5a}
\end{eqnarray}
where  $\xi(t)=\lim_{N\rightarrow \infty} \xi_N(t)$. Thus, the
$N$-fluctuator RTP, with the same switching rates, $\gamma$, and
the amplitudes, $\sigma^2/N$,  for a finite number, $N$, is an
approximation of a Gaussian Markovian process.

\section{Stochastic differential equations}

We consider a system of first-order stochastic differential equations
\begin{align}
\frac{d}{dt} {\mathbf x} (t) = \hat{A}(t){\mathbf x} (t) + \xi(t) \hat{B}(t){\mathbf x}
(t) , \;  {\mathbf x} (0)
= {\mathbf x}_0,
\label{A6}
\end{align}
where  $\xi(t)$  describes ERP, so that
$\langle  {\xi}(t)\xi(t') \rangle = \chi(t-t')$ $( t \geq t')$ and
\begin{align}
\chi(|t-t'|) =  \int dw(\sigma,\gamma) \sigma^2 e^{-2\gamma|t-t'|}.
\label{ERP2}
\end{align}
In what follows we study two approximations leading to a closed system of differential equations
for averaged variables: (i) The effective fluctuator  approximation and (ii) the Gaussian
approximation.

\subsection{Gaussian approximation}

In the interaction picture, we introduce the new variable $\tilde{\mathbf x}(t)=
U^{-1}(t){\mathbf x}(t)$, where
\begin{align}
  U(t) = T\Big(e^{\int\limits_{0}^{t}\hat {A}(t' )dt'}\Big) \tilde{ \mathbf x}(t),
\end{align}
with a T-ordered exponential on the r.h.s. For $\tilde{ \mathbf x}(t)$, Eq. (\ref{A6}) takes the
form
\begin{align}
\frac{d}{dt} \tilde{ \mathbf x} (t) = i \xi(t) \hat{\tilde B}(t)\tilde{\mathbf x} (t) , \;
\tilde{\mathbf x} (0)
= {\mathbf x}_0,
\label{A8}
\end{align}
where we set $ i\hat{\tilde B}(t)= U^{-1}(t)\hat{B}(t)U(t)$. Eq. (\ref{A8}) can be recast as
\begin{align}
\frac{d}{dt} \tilde{ \mathbf x} (t) = i \hat{\tilde B}(t)\xi(t)\tilde{\mathbf x} (0) -
\hat{\tilde B}(t)\int\limits_{0}^{t}\xi(t) \xi(t') \hat{\tilde B}(t'),
\tilde{\mathbf x} (t') d t'.
\label{A8a}
\end{align}
After averaging  over ERP, we obtain the following integro-differential equation
\begin{align}
\frac{d}{dt} \langle \tilde{ \mathbf x} (t)\rangle =  -\hat{\tilde
B}(t)\int\limits_{0}^{t}\chi(t
-t') \hat{\tilde B}(t')
\langle \tilde{\mathbf x} (t')\rangle d t'.
\label{A9}
\end{align}

For practical purposes, Eq. (\ref{A9}) is not very useful. However for some reasonable
assumptions, it can be simplified.
First, employing (\ref{A9}) one can write
\begin{align}
\langle \tilde{ \mathbf x} (t')\rangle  = \langle \tilde{ \mathbf x} (t)\rangle - \int^{t'}_t
dt'\hat{\tilde B}(t')\int\limits_{0}^{t''}\chi(t' -t'') \hat{\tilde B}(t'')\langle
\tilde{\mathbf
x} (t')\rangle d{t''}.
\label{A10}
\end{align}
Then, inserting (\ref{A10}) into Eq. (\ref{A9}) we obtain
\begin{align}
\frac{d}{dt} \langle \tilde{ \mathbf x} (t)\rangle = - \hat{\tilde
B}(t)\int\limits_{0}^{t}\chi(t
-t') \hat{\tilde B}(t')
 d t' \langle \tilde{\mathbf x} (t)\rangle + {\mathcal O}(\parallel\hat{\tilde B}\parallel^4).
\label{A9cr}
\end{align}
Considering the last term at the r.h.s. of Eq. (\ref{A6}) as a perturbation, one can approximate
Eq. (\ref{A9}) as follows:
\begin{align}
\frac{d}{dt} \langle \tilde{ \mathbf x} (t)\rangle = - \hat{\tilde
B}(t)\int\limits_{0}^{t}\chi(t
-t') \hat{\tilde B}(t')
 d t' \langle \tilde{\mathbf x} (t)\rangle.
\label{A9b}
\end{align}
Its formal solution can be written as
\begin{align}
 &\langle \tilde{ \mathbf x} (t)\rangle = T\Big\{\exp\Big(-\int\limits_{0}^{t} \hat{\tilde
 B}(t')
 dt'\int\limits_{0}^{t'}\chi(t' -t'') \hat{\tilde B}(t'')d t''\Big)\Big\}\langle \tilde{\mathbf
 x} (0)\rangle \nonumber \\
 & =  T\Big\{\exp\Big(-\frac{1}{2}\int\limits_{0}^{t}\int\limits_{0}^{t} \hat{\tilde B}(t')
 \chi(t' -t'') \hat{\tilde B}(t'')dt'd t''\Big)\Big\}\langle \tilde{\mathbf x} (0)\rangle.
\label{A9a}
\end{align}
As can be seen, it has the form of solution for the Gaussian random process \cite{KV1,KV2,KV3}.

Returning to $ \langle {\mathbf x} (t) \rangle$, we obtain the first-order differential equation
\begin{align}
\frac{d}{dt} \langle {\mathbf x} (t) \rangle = \hat{A}(t)\langle {\mathbf x} (t) \rangle
+ \hat{B}(t) \hat{ V}(t) \langle {\mathbf x} (t)\rangle,
\label{A11}
\end{align}
where
\begin{align}
\hat{V}(t) = \int_0^t dt'\chi(t-t') U(t) \hat{\tilde B}(t')  U^{-1}(t).
\end{align}

As an illustrative example, let us consider the following stochastic differential equation:
\begin{align}
\frac{d}{dt} { x} (t) = i{A}{ x} (t) + i{v}\xi(t) { x} (t) ,
\end{align}
with $A$ and $v$ being  $\rm const$. Its solution can be written as follows:
\begin{align}
 \langle { x} (t) \rangle = e^{i\varphi_0(t)}   \langle  e^{i\varphi(t)} \rangle  \langle { x}
 (0) \rangle,
 \label{A13}
\end{align}
where $\varphi_0(t) = A t$ is the regular part, and $\varphi(t) = v\int\limits_{0}^{t}\chi(t
-t')
d t'$ is the stochastic phase accumulated at time $t$.

In the Gaussian approximation, we find that the average $\langle { x} (t) \rangle$ satisfies the
differential equation
\begin{align}
\frac{d}{dt} \langle { x} (t) \rangle = i{A}\langle { x} (t) \rangle - {v^
2}\Big(\int\limits_{0}^{t}\chi(t -t')
 d t'\Big) \langle { x} (t)\rangle.
\label{A12}
\end{align}
The solution of  Eq.(\ref{A12}) is given by
\begin{align}
\langle { x} (t) \rangle= \langle  e^{i\varphi_0(t)} \rangle e^{-\kappa(t)} \langle
e^{i\varphi(0)} \rangle,
\end{align}
where
\begin{align}
\kappa(t) = v^2  \int\limits_{0}^{t}dt'\int\limits_{0}^{t'}\chi(t -t') d t''
 = \frac{1}{2} v^2  \int\limits_{0}^{t}\int\limits_{0}^{t}\chi(|t -t'|) dt'd t''  = \frac{1}{2}
 \langle \varphi^2(t) \rangle.
\end{align}
From here and (\ref{A13}), it follows that the decay law for $ \langle  e^{i\varphi(t)} \rangle$
is the Gaussian,
\begin{align}
 \langle  e^{i\varphi(t)} \rangle =   e^{-\langle \varphi^2(t) \rangle/2}.
\end{align}
This agrees with the general conclusions made in this section.

\subsection{Two-effective-fluctuator approximation}

Averaging Eq. (\ref{A6}) over the ERP, we obtain
\begin{eqnarray}
\frac{d}{dt} \langle {\mathbf x} (t) \rangle = \hat{A}(t)\langle
{\mathbf x} (t) \rangle +  \hat{B}(t)\langle {\mathbf  X}_\xi (t)
\rangle, \label{A7}
\end{eqnarray}
where  $ \langle {\mathbf X}_\xi(t) \rangle =\langle \xi (t)
{\mathbf x} (t)\rangle$. Using (\ref{A6}), and taking into account that $\langle {\mathbf
X}_\xi(0) \rangle =0$, we obtain
\begin{eqnarray}
\langle {\mathbf X}_\xi(t) \rangle =\int\limits_{0}^{t}\frac{\chi(t-t')}{\chi(0)}  \hat{A}(t')
\langle {\mathbf  X}_\xi (t') \rangle dt'
+ \int\limits_{0}^{t}
\chi(t-t')\hat{B}(t') \langle {\mathbf  x}(t') \rangle dt'.
\label{A7d}
\end{eqnarray}

Taking the derivative on both sides of Eq. (\ref{A7d}), we obtain
\begin{align}
\frac{d}{dt}\langle {\mathbf X}_\xi(t) \rangle =&
 \hat{A}(t)
\langle {\mathbf  X}_\xi (t) \rangle  + \chi(0) \hat{B}(t) \langle {\mathbf  x}(t) \rangle
\nonumber \\
&+ \frac{1}{\chi(0)}\int\limits_{0}^{t}\frac{\partial\chi(t-t')}{\partial t} \Big (
\hat{A}(t')\langle {\mathbf  X}_\xi (t') \rangle' + \chi(0)\hat{B}(t') \langle {\mathbf  x}(t')
\rangle\Big ) dt'.
\label{A7c}
\end{align}
Finally, we obtain the following closed system of integro-differential equations:
\begin{align}
\frac{d}{dt} \langle {\mathbf x} (t) \rangle = &\hat{A}(t)\langle
{\mathbf x} (t) \rangle +  \hat{B}(t)\langle {\mathbf  X}_\xi (t)
\rangle, \label{A8b} \\
\frac{d}{dt}\langle {\mathbf X}_\xi(t) \rangle =&
 \hat{A}(t) \langle {\mathbf  X}_\xi (t) \rangle
 + \chi(0) \hat{B}(t) \langle {\mathbf  x}(t) \rangle \nonumber \\
&+ \frac{1}{\chi(0)}\int\limits_{0}^{t}\frac{\partial\chi(t-t')}{\partial t} \Big (
\hat{A}(t')\langle {\mathbf  X}_\xi (t') \rangle + \chi(0)\hat{B}(t') \langle {\mathbf  x}(t')
\rangle\Big ) dt'.
\label{A8c}
\end{align}

In this section, we consider the system of Eqs. (\ref{A8b}), (\ref{A8c}) in the approximation
that the ERP can be approximated by a random telegraph process with the correlation function, $
\chi^\ast(|t-t'|)$,
\begin{align}
 \chi(|t-t'|) =  \int dw(\sigma,\gamma) \sigma^2 e^{-2\gamma|t-t'|}
 \approx  \chi^\ast(|t-t'|) = {a^*}^2 e^{-2\gamma^* |t - t'|},
 \label{B2}
\end{align}
where the time-independent parameters, $a^*$ and $\gamma^*$ (the effective amplitude and the
switching rate) are
defined by the following expressions: ${a^*}^2 = \chi(0) $ and
$\gamma^* =- (1/2)\partial \ln\chi(t)/\partial t|_{t=0}$.

To proceed, consider Eq.(\ref{A8c}) rewritten as
\begin{align}
\frac{d}{dt}\langle {\mathbf X}_\xi(t) \rangle =&
 \hat{A}(t) \langle {\mathbf  X}_\xi (t) \rangle  + \chi(0) \hat{B}(t) \langle {\mathbf  x}(t)
 \rangle \nonumber \\
& + \int\limits_{0}^{t}\frac{\partial\ln \chi(t-t')}{\partial t} \chi(t-t')\Big
(\frac{1}{\chi(0)} \hat{A}(t')\langle {\mathbf  X}_\xi (t') \rangle + \hat{B}(t') \langle
{\mathbf  x}(t') \rangle\Big ) dt'.
\label{A8d}
\end{align}
Usually $\ln \chi(t)$ is a slowly-changing function. Then, replacing $\partial
\ln\gamma(t-t')/\partial t $ by its value at time, $t=t'$, one can approximate the integral on
the right side of Eq. (\ref{A8d}) as follows:
\begin{align}
&\int\limits_{0}^{t}\frac{\partial\ln \chi(t-t')}{\partial t} \chi(t-t')\Big (\frac{1}{\chi(0)}
\hat{A}(t')\langle {\mathbf  X}_\xi (t') \rangle' + \hat{B}(t') \langle {\mathbf  x}(t')
\rangle\Big ) dt' \nonumber \\
&\approx \frac{\partial\ln \chi(t-t')}{\partial t}\Big|_{t=t'}\int\limits_{0}^{t} \chi(t-t')\Big
(\frac{1}{\chi(0)} \hat{A}(t')\langle {\mathbf  X}_\xi (t') \rangle + \hat{B}(t') \langle
{\mathbf  x}(t') \rangle\Big ) dt'.
\label{A8e}
\end{align}

Inserting (\ref{A8e}) into (\ref{A8d}), and employing (\ref{A7d}) we obtain
\begin{align}
\frac{d}{dt}\langle {\mathbf X}_\xi(t) \rangle =\hat{A}(t) \langle {\mathbf  X}_\xi (t) \rangle
-2\gamma^*\langle {\mathbf X}_\xi(t) \rangle
 + {a^\ast}^2 \hat{B}(t) \langle {\mathbf  x}(t) \rangle,
 \label{A7A}
\end{align}
where  $\gamma^* =- (1/2)\partial \ln\chi(t)/\partial t|_{t=0}$ and ${a^\ast}^2 = \chi(0) $.
Next, combining (\ref{A7}) and (\ref{A7A}), instead of a system of integro-differential
equations, we obtain a closed system of first-order differential equations
\begin{align}
\label{A14}
\frac{d}{dt} \langle {\mathbf x} (t) \rangle = \hat{A}(t)\langle {\mathbf x} (t) \rangle
 + \hat{B}(t)\langle {\mathbf  X}_\xi (t) \rangle, \quad
  \langle {\mathbf x} (0) \rangle = {\mathbf x} (0 ),  \\
\frac{d}{dt}\langle {\mathbf X}_\xi(t) \rangle + 2\gamma^*\langle {\mathbf X}_\xi(t) \rangle =
\hat{A}(t) \langle {\mathbf  X}_\xi (t) \rangle + {a^\ast}^2 \hat{B}(t) \langle {\mathbf  x}(t)
\rangle, \quad \langle {\mathbf X}_\xi(0) \rangle =0.
\label{A14a}
\end{align}
This system of differential equations describes RTP with the amplitude $a^\ast$, switching rate
$\gamma^*$ and the correlation function \cite{KV3}
\begin{align}
 \chi^\ast(|t-t'|) = {a^*}^2 e^{-2\gamma^* |t - t'|}.
 \label{A12a}
\end{align}

In Fig. \ref{CHI_5}, we compare the exact correlation functions,
\begin{align}
\chi_{1}(\tau) = \sigma^2_{1} A_1(E_1(2\gamma_m \tau) - E_1(2\gamma_c \tau) ), \\
\chi_{2}(\tau) = \sigma^2_{2} A_2 \bigg (\frac{ E_2(2\gamma_c \tau)}{\gamma_c} - \frac{
E_2(2\gamma_0 \tau)}{\gamma_0} \bigg),
\end{align}
with their approximated expressions, $\chi_n
\approx \sigma^2_n\exp(-2\gamma_n t)$ given by (\ref{ccc}). The choice of
parameters, $\gamma_m$ and $\gamma_c$, was motivated by  the range
of frequencies for $1/f$ noise. The parameter, $\gamma_0$, was
chosen to better fit both exact and approximate correlation
functions. Note, that the correlation function in (\ref{ccc}) which
describes the low frequency noise, $\chi_1$, is not very sensitive to
variations of the parameter, $\gamma_m$. Further, when fitting the experimental data, the
parameters, $\gamma_m$ and $\gamma_c$, were essentially the same as in Fig. \ref{CHI_5}.  As can
be seen, the approximation (\ref{B2}) describes the behavior of the exact correlation functions
reasonably well for the region of parameters which we use.
\begin{figure}[tbh]
\begin{center}
\scalebox{0.35}{\includegraphics{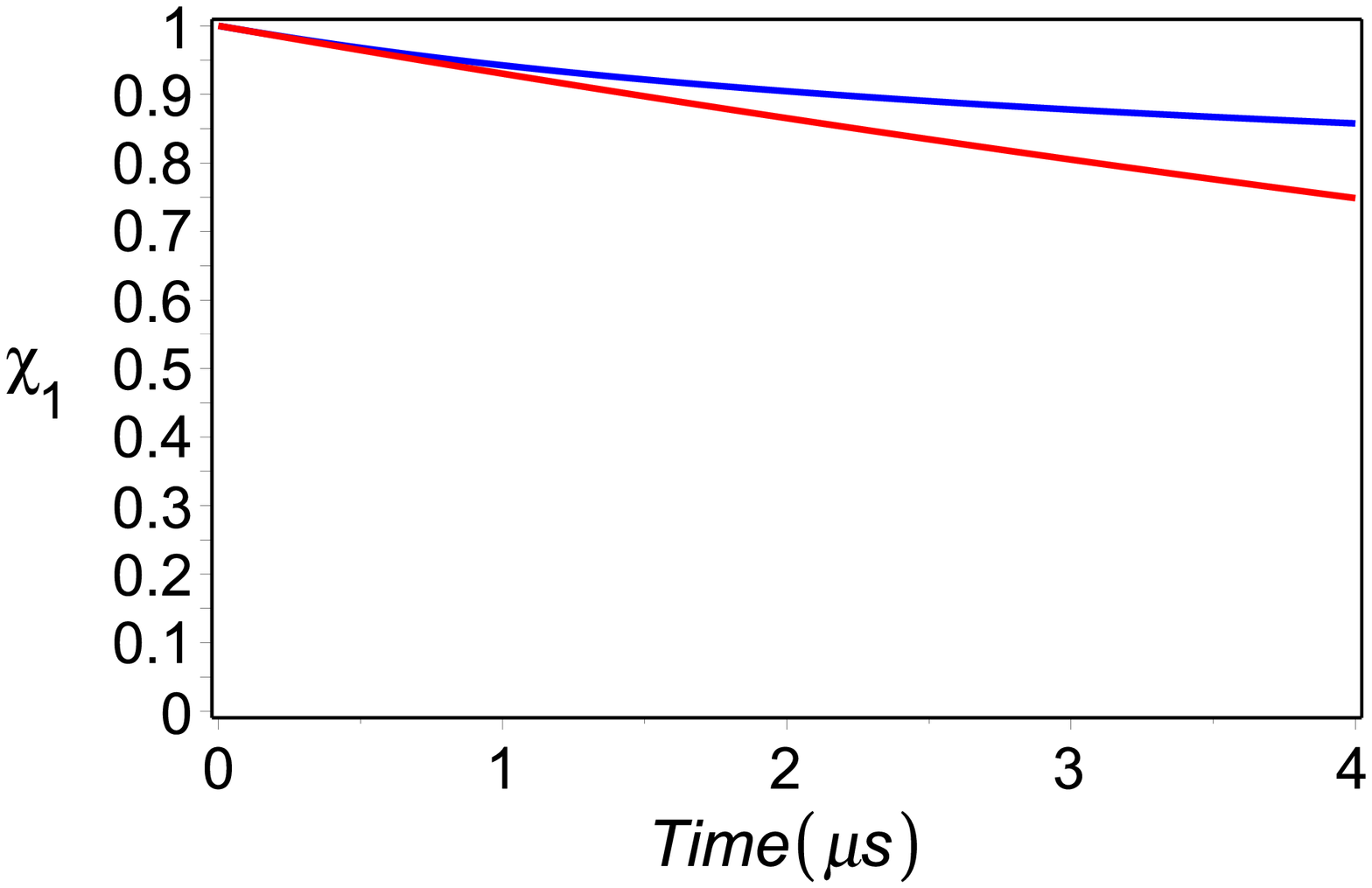}}
\scalebox{0.35}{\includegraphics{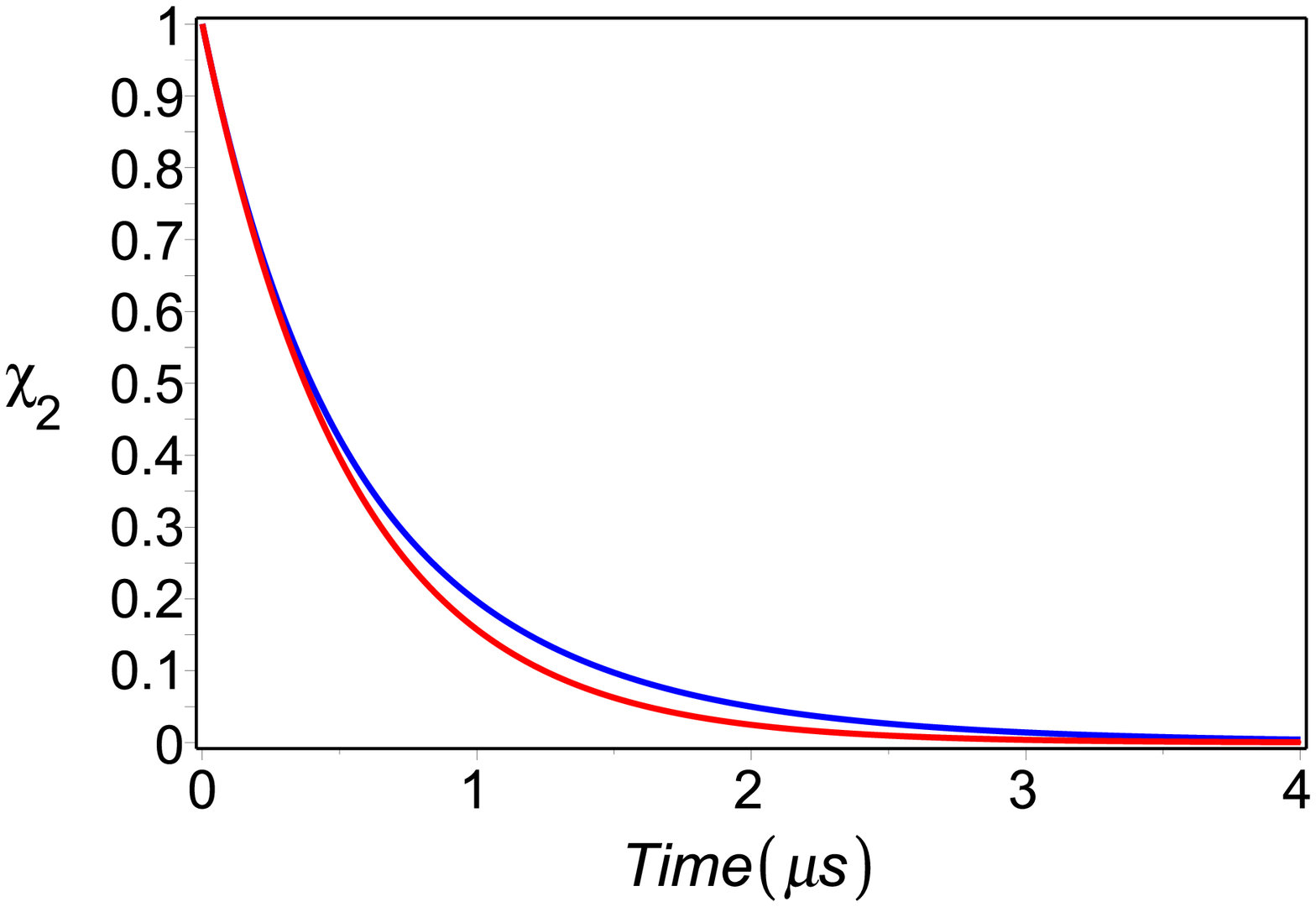}}
\end{center}
\caption{Correlation functions, $\chi_n(t)$, (blue line)  and exponential correlation functions,
$\chi_n(t) = \exp(-2\gamma_n t)$, (red line). Upper panel: Low-frequency noise defined by
$\chi_1(t)$  ($\gamma_m = 0.5 \;\rm  s^{-1} $,  $\gamma_c = 0.5 \;\rm \mu s^{-1} $). There is
good agreement up to $\sim 2 \mu s$. Bottom panel: High- frequency noise  defined by $\chi_2(t)$
($\gamma_c = 0.5 \;\rm \mu s^{-1} $,  $\gamma_0 = 2 \;\rm \mu s^{-1} $). In all cases
$\sigma_n^2
= 1$.} \label{CHI_5}
\end{figure}

The system of Eqs. (\ref{A14}), (\ref{A14a}) approximately describes an
ERP by RTP defined by a single fluctuator.
Below, we will describe a model with two effective (low- and high-frequency) fluctuators. The
advantage of this approach is that we calculate in a straightforward way the coefficients $a^*$
and $\gamma^*$.

\subsubsection*{Two-effective-fluctuators model}

Let us  consider the same system of first-order stochastic differential equations as above
\begin{align}
\frac{d}{dt} {\mathbf x} (t) = \hat{A}(t){\mathbf x} (t) + \xi(t) \hat{B}(t){\mathbf x} (t) , \;
{\mathbf x} (0)
= {\mathbf x}_0,
\label{A15}
\end{align}
with the RTP described by two uncorrelated fluctuators, $\zeta_1(t)$ and 2$\zeta_2(t)$, so that
$\xi(t) =  \zeta_2(t) + \zeta_2(t)$, and
\begin{align}
&\langle  {\zeta}_i(t) \rangle = 0 , \\
&\langle  {\zeta}_i(t) {\zeta}_j(t') \rangle = \delta_{ij}a^2_i
e^{-2\gamma_i|t-t'|} , \quad i=1,2 .
\label{C6r}
\end{align}
We set $\langle {\mathbf  X}_1 (t) \rangle = \langle \zeta_1(t) {\mathbf  x} (t) \rangle $,
 $\langle {\mathbf  X}_2 (t) \rangle = \langle \zeta_2(t) {\mathbf  x} (t) \rangle $ and
 $\langle {\mathbf  X}_{12} (t) \rangle = \langle \zeta_1(t)  \zeta_2(t) {\mathbf  x} (t)
 \rangle
 $. Applying the formulae of differentiation for an RTP \cite{KV1,KV2,KV3}, we obtain the
 following system of differential equations for averaged variables:
\begin{align}
\frac{d}{dt} \langle {\mathbf x} (t) \rangle = \hat{A}(t)\langle {\mathbf x} (t) \rangle
+  \hat{B}(t)(\langle {\mathbf  X}_1 (t) \rangle + \langle {\mathbf  X}_2 (t) \rangle),
\nonumber
\\
\frac{d}{dt}\langle {\mathbf X}_1(t) \rangle = -2\gamma_1\langle {\mathbf X}_1(t) \rangle +
\hat{A}(t) \langle {\mathbf  X}_1 (t) \rangle
 + \hat{B}(t) (\langle {\mathbf  X}_{12}(t) \rangle +a^2_1\langle {\mathbf  x}(t) \rangle),
 \nonumber \\
 \frac{d}{dt}\langle {\mathbf X}_2(t) \rangle = -2\gamma_2\langle {\mathbf X}_1(t) \rangle +
 \hat{A}(t) \langle {\mathbf  X}_2 (t) \rangle
 + \hat{B}(t) (\langle {\mathbf  X}_{12}(t) \rangle +a^2_2\langle {\mathbf  x}(t) \rangle),
 \nonumber \\
 \frac{d}{dt}\langle {\mathbf X}_{12}(t) \rangle = -2(\gamma_1+\gamma_2)\langle {\mathbf
 X}_{12}(t) \rangle + \hat{A}(t) \langle {\mathbf  X}_{12} (t) \rangle
 + \hat{B}(t) (a^2_2\langle {\mathbf  X}_{1}(t) \rangle + a^2_1\langle {\mathbf  X}_2(t)
 \rangle).
\label{A16}
\end{align}

\section{Properties of the correlation functions}

We consider a family of random variables and distributions, $\{\xi_n(t),dw_n(\sigma,\gamma)\}$,
in which each $\xi_n(t)$ describes an independent ERP: $\langle \xi_m(t)\xi_n(t')\rangle =0$
$(m\neq n)$. Then, the total correlation function is a sum of the partial correlation functions
and $\chi(|t-t'|)=\sum_n\chi_{n}(|t-t'|)$, can be written as
\begin{eqnarray}
  \chi(|t-t'|) = \sum_n  \iint dw_n(\sigma,\gamma)  \sigma^2   e^{-2\gamma|t-t'|}.
 \label{Ccn}
\end{eqnarray}

We define the distribution function, $dw_n(\sigma,\gamma)$, as
\begin{eqnarray}
dw_n(\sigma,\gamma) =\delta(\sigma-  \sigma_n){\mathcal P}_n(\gamma) d\sigma d\gamma,
\end{eqnarray}
where, $\sigma_n$, is a some typical value of the amplitude, and
\begin{eqnarray}\label{CNeq11}
&{\mathcal P}_n(\gamma)d\gamma=A_n\Theta(\gamma_{c_n} - \gamma)\Theta
(\gamma - \gamma_{m_n})\displaystyle\frac{d\gamma}{\gamma^n}, \quad n=1,2,\dots,
 \end{eqnarray}
here, $\Theta(x)$, denotes the step-function,  $\gamma_{m_n}$ and $\gamma_{c_n}$ are the lower
and upper switching rates, respectively. The normalization constant given by,
\begin{eqnarray}
A_n =\left \{  \begin{array}{ll}
\displaystyle \frac{1}{\ln(\gamma_{c_1}/\gamma_{m_1})}, & n=1\\
&\\
\displaystyle \frac{(n-1)\gamma^{n-1}_{m_n}}{(1- \gamma^{n-1}_{m_n}/
\gamma^{n-1}_{c_n})}, & n \neq 1
\end{array}
\right .,
\end{eqnarray}
is obtained from the normalization condition, $\int dw_n(\sigma,\gamma) =1$.

Inserting (\ref{CNeq11}) into  (\ref{Ccn}), we obtain,
\begin{eqnarray}
\chi_{n}(|t-t'|)  =  { \sigma}_n^2\iint{\mathcal P}_n(\gamma)d\gamma    e^{-2\gamma|t-t'|} .
\label{CR1}
\end{eqnarray}
From  (\ref{CR1}) it follows that $\sigma^2= \chi(0)$, and straightforward computation  yields,
\begin{eqnarray}\label{Chi_1}
\chi_{n}(\tau) = {\sigma^2_{n} A_n}\Bigg( \frac{E_n(2 \gamma_{m_n}\tau)}{ \gamma_{m_n}^{n-1}}-
\frac{E_n(2 \gamma_{c_n}\tau)}{ \gamma_{c_n}^{n-1}} \Bigg),
\end{eqnarray}
where $E_n(z)$ denotes  the Exponential integral \cite{abr}.

It is convenient to describe each noise source by its spectral density,
\begin{eqnarray}
S_{n}(\omega) = \frac{1}{\pi}\int\limits_{0}^{\infty}
\chi_n(\tau) \cos(\omega \tau )d\tau ,
\label{Sf1}
\end{eqnarray}
and, as it can be easily seen, $\sigma^2_{n}=2\int\limits_{0}^{\infty}S_n(\omega)d\omega$.
Employing Eqs. (\ref{Ccn}) and \ref{Sf1}), one can obtain the following integral representation
for $S_n(\omega)$:
\begin{eqnarray}
S_n(\omega) =\frac{1}{\pi}\int \frac{ 2\gamma \sigma^2}{4\gamma^2 +
\omega^2}\,dw_n(\sigma,\gamma),
\end{eqnarray}
where
\begin{align}
  S_L(\Omega) = \frac{1}{\pi}\frac{2\gamma\sigma^2 }{4\gamma^2 +
\omega^2}
\end{align}
is the Lorentzian spectral density of the fluctuator with the amplitude $\sigma$ and switching
rate $\gamma$ \cite{BGA}.

Performing the integration in Eq. (\ref{Sf1}), we obtain for $n>2$
\begin{align}
S_n(\omega) =  \frac{1}{\pi}\sigma^2_{n} A_n 2 ^{n-1}\sum^{[(n+1)/2]}_{k=1}
\frac{(-1)^{k+1}}{(n-2k)\omega^{2k}}\bigg(\frac{1}{b_n^{n-2k}}- \frac{1}{c_n^{n-2k}}\bigg) \\
+ \frac{1}{\pi\omega^{n}} A_n \sigma^2_{n}2 ^{n-1}\left \{
\begin{array}{ll}
\displaystyle  \frac{1}{2}\ln\bigg(\frac{1 + ({\omega}/{b_n})^2}{1 + ({\omega}/{c_n})^2}\bigg)
,&
n=2p, \\
&\\
\displaystyle  \arctan \Big(\frac{\omega}{b_n}\Big)- \arctan\Big(\frac{\omega}{c_n}\Big), &
n=2p+
1,
\end{array}
 \right .
 \label{SF2a}
\end{align}
where $b_n= 2\gamma_{m_{n}}$ and  $c_n= 2\gamma_{c_{n}}$. For $n=1,2$, the computation yields
\begin{align}
&S_{1}(\omega) =  \frac{\sigma^2_{1} A_1}{\pi\omega}\bigg(\arctan \Big(\frac{\omega}{b_1}\Big)-
\arctan\Big(\frac{\omega}{c_1}\Big)\bigg), \\
&S_{2}(\omega) = \frac{\sigma^2_{2} A_2}{\pi\omega^2}\ln\bigg(\frac{1 + ({\omega}/{b_2})^2}{1 +
({\omega}/{c_2})^2}\bigg)
\end{align}

We impose on the distribution functions ${\mathcal P}_1(\gamma)$ and ${\mathcal P}_2(\gamma)$
boundary conditions at the point $\gamma = \gamma_c$, so that $\gamma_{m _2}= \gamma_{c _1}$.
Further, we denote $\gamma_m =\gamma_{m _1}$, $\gamma_c =\gamma_{c _1} $, and $\gamma_0
=\gamma_{c _2} $  $(\gamma_m <\gamma_c <\gamma_0)$. Using these notations, we obtain
\begin{align}\label{ASF1}
&S_{1}(\omega) = \frac{\sigma^2_{1} A_1}{\pi\omega}\bigg( \arctan
\Big(\frac{\omega}{2\gamma_m}\Big)- \arctan\Big(\frac{\omega}{2\gamma_c}\Big)\bigg), \\
&S_{2}(\omega) = \frac{\sigma^2_{2} A_2}{\pi\omega^2}\ln\bigg(\frac{1 + \omega^2/4\gamma_c^2}{1
+
\omega^2/4\gamma_0^2}\bigg),
\label{ASF2}
\end{align}

This yields the following asymptotic behavior of $S_1(\omega)$ and $S_2(\omega)$:
\begin{eqnarray}
S_{1}(\omega) \approx \left \{\begin{array}{ll}
\displaystyle\frac{\sigma^2_{1} }{2\pi \gamma_m \ln(\gamma_c/\gamma_m) }\bigg(1-
\frac{\gamma_m}{\gamma_c}\bigg),&   \omega \ll2\gamma_m ,\\
\\
\displaystyle\frac{\sigma^2_{1} }{2  \omega\ln(\gamma_c/\gamma_m)},& 2\gamma_m \ll  \omega
\ll2\gamma_c, \\
\\
\displaystyle\frac{2\sigma^2_{1} \gamma_c (1-\gamma_m/\gamma_c)}{\pi \omega^2
\ln(\gamma_c/\gamma_m)},&   \omega \gg 2\gamma_c,
\end{array}
\right.
\end{eqnarray}
and
\begin{eqnarray}
S_{2}(\omega) \approx \left \{\begin{array}{ll}
\displaystyle\frac{\sigma^2_{2}  }{4\pi  \gamma_c }\bigg(1+\frac{\gamma_c}{\gamma_0}\bigg),&
\omega \ll2\gamma_c < 2\gamma_0,\\
\\
\displaystyle\frac{2\sigma^2_{2} \gamma_c }{\pi(1- \gamma_c/\gamma_0)\omega^2}\ln\bigg(\frac{
\omega}{2\gamma_c}\bigg),& 2\gamma_c \ll  \omega \ll2\gamma_0 ,\\
\\
\displaystyle\frac{2 \sigma^2_{2} \gamma_c}{\pi (1- \gamma_c/\gamma_0)\omega^2}\ln\bigg(\frac{
\gamma_0}{\gamma_c}\bigg),&   \omega \gg 2\gamma_0.
\end{array}
\right .
\end{eqnarray}

\begin{eqnarray}
\frac {S_{2}(\omega)}{S_{1}(\omega)}\approx \left \{\begin{array}{ll}
\displaystyle\frac{\sigma^2_{2}  }{ \sigma^2_{1}}\frac{ \gamma_m  \ln(\gamma_c/\gamma_m)}{2
\gamma_c}\bigg(1+\frac{\gamma_c}{\gamma_0}\bigg),&   \omega \approx 0,\\
\\
\displaystyle \frac{\sigma^2_{2}  }{ \sigma^2_{1}} \frac{ \ln(\gamma_c/\gamma_m) }{2(1-
\gamma_c/\gamma_0)}\ln\bigg(\frac{1 + \omega^2/4\gamma_c^2}{1 + \omega^2/4\gamma_0^2}\bigg),&
\omega\gtrsim 2\gamma_c .
\end{array}
\right .
\end{eqnarray}
From Eqs. (\ref{SF1}) and (\ref{SF2}), it follows that in the interval, $\gamma_m <  \omega
<\gamma_c$, the spectral density $S_{1}(\omega)$  describes  $1/f$ noise. Indeed, in this
interval $S_{1}(\omega) \approx A/\omega$, where $A= \sigma^2_{1}/(2\ln(\gamma_c/\gamma_m)) $.
For $S_2(\omega)$ we obtain the following asymptotic behavior: $S_2(\omega) \sim 1/\omega^2$
($\omega \gg \omega_c$). Thus, asymptotically $S_2(\omega)$ yields the Lorentzian spectrum.

Writing the spectral density for  $1/f$ noise as
$S_{1/f} (\omega)=A\Theta(\omega_c - \omega)\Theta(\omega -\omega_m)/\omega $, where $\omega_c$
and $\omega_m$ are ultraviolet and infrared cutoff, respectively, we obtain
\begin{eqnarray}
 \sigma^2_{1}=2 \int_{0}^{\infty}S_{1} (\omega)d\omega \approx 2\int_{0}^{\infty}S_{1/f}
 (\omega)d\omega =2A
 \ln(\omega_c/\omega_m)=\sigma^2_{1}\frac{\ln(\omega_c/\omega_m)}{\ln(\gamma_c/\gamma_m) }.
\end{eqnarray}
From here it follows:  $\gamma_c/\gamma_m \approx \omega_c/\omega_m$. Thus,
$\gamma_m$ and $\gamma_c$  are
related to the infrared and ultraviolet frequency cutoff, respectively. Further we assume
$\omega_c = 2\gamma_c$ and $ \omega_m = 2\gamma_m $.

As can be seen from Eq. (\ref{SF2a}), our model covers various asymptotic aspects of the
spectral
density, $S(\omega)= \sum_nS_n(\omega)$, including $1/f$ noise and the Lorentzian spectrum as
some particular cases. This allows us to include into consideration the more complicated
behaviors of the spectral density. \\

{\em Estimates of correlation times for superconducting qubits.} Following \cite{KV1,KV3}, we
define the correlation time related to $\chi_n(\tau)$ as
\begin{eqnarray}
\tau_n   =\frac{1}{\chi_{n}(0)} \int^\infty_0 \chi_{n}(\tau) d\tau .
\label{Corr1}
\end{eqnarray}
From here, employing Eq. (\ref{Chi_1}), we obtain
\begin{eqnarray}
\tau_n = \left \{
\begin{array}{ll}
\displaystyle \frac{1- b_1/c_1}{b_1\ln(c_1/b_1)},& n=1,\\
&\\
\displaystyle \frac{(n-1)(1- (b_n/c_n)^n)}{n b_n (1- (b_n/c_n)^{n-1} )} , & n\neq 1.
\end{array}
 \right .
\end{eqnarray}
For  $b_n \ll c_n$, this yields
\begin{align}
\tau_n \approx \left \{
\begin{array}{ll}
\displaystyle \frac{1}{b_1\ln(c_1/b_1)},& n=1,\\
&\\
\displaystyle \frac{n-1}{n b_n } , & n\neq 1.
\end{array}
 \right .
\end{align}

Using Eq. (\ref{Corr1}), we calculate  the correlation time of $1/f$ noise to be
\begin{eqnarray}
\tau_{1}  = \frac{1- \gamma_m/\gamma_c}{2\gamma_m \ln(\gamma_c/\gamma_m)}.
 \label{t1}
\end{eqnarray}
For $\gamma_m  \ll \gamma_c$, this yields
\begin{eqnarray}
\tau_{1}  \approx \frac{1}{2\gamma_m \ln(\gamma_c/\gamma_m)} .
 \label{T1}
\end{eqnarray}
Computation of the correlation time $\tau_{2}$  yields
\begin{eqnarray}
\tau_{2}   = \frac{1}{4\gamma_c} \bigg(1+ \frac{\gamma_c}{\gamma_0}\Big).
\label{t1a}
\end{eqnarray}
From Eqs. (\ref{t1}) and (\ref{t1a}) we obtain
\begin{eqnarray}
\frac{\tau_{2}}{\tau_{1}}   \lesssim \frac{\gamma_m}{2\gamma_c}\ln(\gamma_c/\gamma_m).
\label{t2}
\end{eqnarray}

For superconducting qubits various experiments demonstrate that the
frequency interval of $1/f$ noise is $ f\sim(1\rm Hz - 1\rm
MHz)$ \cite{ICJM}. Substituting  $2\gamma_m =1\rm s^{-1}$ and $2\gamma_c =1\rm \mu s^{-1}$ into
(\ref{T1}), we obtain an estimate of the effective correlation times as $\tau_{1} \sim 0.01 s$.
The experimental data on the ultraviolet cutoff of the spectral density are unknown, so
$\gamma_0$ is unknown parameter. Supposing $\gamma_0 \gg \gamma_c$, one can estimate the
effective correlation time as $\tau_{2}  \sim 1/(4\gamma_c)$. Once again, assuming that
$\gamma_c
\sim 0.5\rm \mu s^{-1}$,  we obtain $\tau_{2} \sim 0.5\mu s $. So, the fluctuations due to
$\xi_{2}(t) $ have a shorter correlation times than fluctuations related to $1/f$ noise,
$\tau_{2} \ll\tau_{1}$. Thus, indeed, the SF produce mainly noise with the spectrum $\sim
1/\omega$, and the FF lead to the spectrum
$\sim 1/\omega^2$.

\subsection{Free induction signal decay}

For a superconducting qubit in the Gaussian approximation, free induction signal decay is defined
by $\langle e^{i\varphi(t)}\rangle = e^{-(1/2)\langle \varphi^2(t)\rangle}$,
where $\varphi(t) = D_{\lambda,z}\int\limits_{0}^{t}\delta\lambda(t')dt'$ is  the random phase
accumulated at time $t$, and
\begin{align}
 \langle \varphi^2(t) \rangle =
 D^2_{\lambda,z}\int\limits_{0}^{t}\int\limits_{0}^{t}\chi_\lambda(|t' -t''|) dt'd t''.
\end{align}
The correlation function, $\chi_\lambda(\tau)$, of the ERP defined as $\delta\lambda(t)= \sum_n
\xi_n(t)$, can be written as the sum of the partial correlation functions, $\chi_\lambda(\tau) =
\sum_{n}\chi_{n}(\tau)$, and the overall accumulated random phase, $\varphi(t)$, is given by
$\varphi(t) =  \sum_{n} D_{\lambda,z}\int\limits_{0}^{t} \xi_n(t')dt'$. From this we obtain
$\langle \varphi^2(t) \rangle = \sum_{n} \langle \varphi_{n}^2(t) \rangle $, where
\begin{align}
 \langle \varphi_{n}^2(t) \rangle =
 D^2_{\lambda,z}\int\limits_{0}^{t}\int\limits_{0}^{t}\chi_{n}(|t' -t''|) dt'd t'' .
\end{align}
Computation of $ \langle \varphi_{n}^2(t) \rangle$ yields
\begin{align}
 \langle \varphi_{n}^2(t) \rangle = 2^nD^2_{\lambda,z}  \sigma^2_{n} A_n \Bigg(
 \frac{E_{n+2}(b_n
 t)}{b^{n+1}_n}-  \frac{E_{n+2}(c_n t)}{c^{n+1}_n} +\frac{1}{n+1}\Big(\frac{1}{c^{n+1}_n}
 -\frac{1}{b^{n+1}_n}\Big)  + \frac{t}{n}\Big(\frac{1}{b^{n}_n} -\frac{1}{c^{n}_n}\Big)\Bigg)
  \label{C7}
\end{align}

\subsection{Echo decay}

In echo experiments, the total phase, $\psi(t)$, is defined as the difference between two free
evolutions \cite{BGA,ICJM},
\begin{align}
\psi(t) =  D_{\lambda,z}\int\limits_{0}^{t/2}\delta\lambda(t')dt' -
D_{\lambda,z}\int\limits_{t/2}^{t}\delta\lambda(t')dt'.
\end{align}
In the Gaussian approximation, one obtains $\langle e^{i\psi(t)}\rangle = e^{-(1/2)\langle
\psi^2(t)\rangle}$, where
\begin{align}\label{C9}
\langle \psi^2(t)\rangle =  D^2_{\lambda,z} \Bigg (\int\limits_{0}^{t}\int\limits_{0}^{t}dt' d
t'' \chi_\lambda(|t' -t''|)
 - 4\int\limits_{0}^{t/2} dt' \int\limits_{t/2}^{t} d t'' \chi_\lambda(|t' -t''|) \Bigg ).
\end{align}
Inserting $\chi_\lambda(|t' -t''|) = \sum_{n}\chi_{n}(|t' -t''|)$ into Eq. (\ref{C9}), we obtain
$\langle \psi^2(t) \rangle = \sum\limits_{n} \langle \psi_{n}^2(t) \rangle $, where
 \begin{align}
\langle \psi_{n}^2(t) \rangle =  D^2_{\lambda,z} \Bigg
(\int\limits_{0}^{t}\int\limits_{0}^{t}dt'
d t'' \chi_{n} (|t' -t''|)
 - 4\int\limits_{0}^{t/2} dt' \int\limits_{t/2}^{t} d t'' \chi_{n} (|t' -t''|) \Bigg ).
\end{align}
Computation yields
\begin{align}
 \langle \psi_{n}^2(t) \rangle =2^n D^2_{\lambda,z}  \sigma^2_{n} A_n \Bigg( 4\frac{E_{n+2}(b_n
 t/2)}{b^{n+1}_n} - 4\frac{E_{n+2}(c_n t/2)}{c^{n+1}_n} + \frac{E_{n+2}(c_n t)}{c^{n+1}_n}
 \nonumber \\
  -  \frac{E_{n+2}(b_n t)}{b^{n+1}_n}
 +\frac{3}{n+1}\Big(\frac{1}{c^{n+1}_n} -\frac{1}{b^{n+1}_n}\Big)  +
 \frac{t}{n}\Big(\frac{1}{b^{n}_n} -\frac{1}{c^{n}_n}\Big)\Bigg)
 \label{C8}
\end{align}
\end{widetext}

\end{document}